\pdfoutput=1
\documentclass[12pt,a4paper]{article}
\usepackage{ifthen} %
\newboolean{pdflatex}
\setboolean{pdflatex}{true} %

\newboolean{articletitles}
\setboolean{articletitles}{true} %

\newboolean{uprightparticles}
\setboolean{uprightparticles}{false} %

\newboolean{inbibliography}
\setboolean{inbibliography}{false} %

\newboolean{wordcount}
\setboolean{wordcount}{false} %

\ifthenelse{\boolean{wordcount}}%
{\usepackage{bibentry} %
 \usepackage{mciteplus} %
 \usepackage{comment} %
 \excludecomment{acknowledgments} %
 \def\maketitle{} %
}{
}

\def\paperauthors{LHCb collaboration} %
\def\paperasciititle{Angular analysis of D0->pi+pi-mu+mu- and D0->K+K-mu+mu- decays and search for CP violation} %
\def\papertitle{Angular analysis\\
of \Dppmm\\
and \Dkkmm decays\\
and search for \CP violation} %
\def\paperkeywords{{High Energy Physics}, {LHCb}} %
\def\papercopyright{\the\year\ CERN for the benefit of the LHCb collaboration} %
\def\paperlicence{CC BY 4.0 licence}
\def\paperlicenceurl{https://creativecommons.org/licenses/by/4.0/}

\usepackage[top=1in, bottom=1.25in, left=1in, right=1in]{geometry}

\usepackage{microtype}
\usepackage{lineno}  %
\usepackage{xspace} %
\usepackage{caption} %

\usepackage{graphicx}  %
\usepackage{color}
\usepackage{colortbl}
\graphicspath{{./figs/}} %
\usepackage{amsmath} %
\usepackage{amssymb}
\usepackage{amsfonts}
\usepackage{upgreek} %

\newcommand*\patchAmsMathEnvironmentForLineno[1]{%
\expandafter\let\csname old#1\expandafter\endcsname\csname #1\endcsname
\expandafter\let\csname oldend#1\expandafter\endcsname\csname
end#1\endcsname
 \renewenvironment{#1}%
   {\linenomath\csname old#1\endcsname}%
   {\csname oldend#1\endcsname\endlinenomath}%
}
\newcommand*\patchBothAmsMathEnvironmentsForLineno[1]{%
  \patchAmsMathEnvironmentForLineno{#1}%
  \patchAmsMathEnvironmentForLineno{#1*}%
}
\AtBeginDocument{%
\patchBothAmsMathEnvironmentsForLineno{equation}%
\patchBothAmsMathEnvironmentsForLineno{align}%
\patchBothAmsMathEnvironmentsForLineno{flalign}%
\patchBothAmsMathEnvironmentsForLineno{alignat}%
\patchBothAmsMathEnvironmentsForLineno{gather}%
\patchBothAmsMathEnvironmentsForLineno{multline}%
\patchBothAmsMathEnvironmentsForLineno{eqnarray}%
}

\usepackage{hyperxmp}

\usepackage[pdftex,
            pdfauthor={\paperauthors},
            pdftitle={\paperasciititle},
            pdfkeywords={\paperkeywords},
            pdfcopyright={Copyright (C) \papercopyright},
            pdflicenseurl={\paperlicenceurl}]{hyperref}
\usepackage[bottom,flushmargin,hang,multiple]{footmisc}

\usepackage[all]{hypcap} %

\usepackage{xspace} 
\usepackage{upgreek}

\def\lhcb   {\mbox{LHCb}\xspace}

\def\MagUp {\mbox{\em Mag\kern -0.05em Up}\xspace}

\ifthenelse{\boolean{uprightparticles}}%
{

 \def\Pmu         {\ensuremath{\upmu}\xspace}

 \def\Ppi         {\ensuremath{\uppi}\xspace}                 
                  
 \def\Prho        {\ensuremath{\uprho}\xspace}

 \def\Pomega      {\ensuremath{\upomega}\xspace}                 

 \def\PDelta      {\ensuremath{\Delta}\xspace}                 
 \def\PXi         {\ensuremath{\Xi}\xspace}                 
 \def\PLambda     {\ensuremath{\Lambda}\xspace}                 
 \def\PSigma      {\ensuremath{\Sigma}\xspace}                 
 \def\POmega      {\ensuremath{\Omega}\xspace}                 
 \def\PUpsilon    {\ensuremath{\Upsilon}\xspace}

 \def\PB      {\ensuremath{\mathrm{B}}\xspace}                 
                  
 \def\PD      {\ensuremath{\mathrm{D}}\xspace}

 \def\PK      {\ensuremath{\mathrm{K}}\xspace}

 \def\Pb      {\ensuremath{\mathrm{b}}\xspace}                 
 \def\Pc      {\ensuremath{\mathrm{c}}\xspace}

 \def\Pi      {\ensuremath{\mathrm{i}}\xspace}

 \def\Pp      {\ensuremath{\mathrm{p}}\xspace}

 \def\Ps      {\ensuremath{\mathrm{s}}\xspace}                 
                  
 \def\Pu      {\ensuremath{\mathrm{u}}\xspace}

 \def\thebaroffset{0.0em}
}
{

 \def\Pmu         {\ensuremath{\mu}\xspace}

 \def\Ppi         {\ensuremath{\pi}\xspace}                 
                  
 \def\Prho        {\ensuremath{\rho}\xspace}

 \def\Pomega      {\ensuremath{\omega}\xspace}                 
 \mathchardef\PDelta="7101
 \mathchardef\PXi="7104
 \mathchardef\PLambda="7103
 \mathchardef\PSigma="7106
 \mathchardef\POmega="710A
 \mathchardef\PUpsilon="7107
                  
 \def\PB      {\ensuremath{B}\xspace}                 
                  
 \def\PD      {\ensuremath{D}\xspace}

 \def\PK      {\ensuremath{K}\xspace}

 \def\Pb      {\ensuremath{b}\xspace}                 
 \def\Pc      {\ensuremath{c}\xspace}

 \def\Pi      {\ensuremath{i}\xspace}

 \def\Pp      {\ensuremath{p}\xspace}

 \def\Ps      {\ensuremath{s}\xspace}                 
                  
 \def\Pu      {\ensuremath{u}\xspace}

 \def\thebaroffset{0.18em}
}
\newcommand{\offsetoverline}[2][\thebaroffset]{\kern #1\overline{\kern -#1 #2}}%

\makeatletter
\ifcase \@ptsize \relax%
  \newcommand{\miniscule}{\@setfontsize\miniscule{4}{5}}%
\or%
  \newcommand{\miniscule}{\@setfontsize\miniscule{5}{6}}%
\or%
  \newcommand{\miniscule}{\@setfontsize\miniscule{5}{6}}%
\fi
\makeatother

\DeclareRobustCommand{\optbar}[1]{\shortstack{{\miniscule (\rule[.5ex]{1.25em}{.18mm})}
  \\ [-.7ex] $#1$}}

\def\mup        {{\ensuremath{\Pmu^+}}\xspace}
\def\mun        {{\ensuremath{\Pmu^-}}\xspace} %

\def\mumu       {{\ensuremath{\Pmu^+\Pmu^-}}\xspace}

\def\uquark    {{\ensuremath{\Pu}}\xspace}

\def\squark    {{\ensuremath{\Ps}}\xspace}

\def\cquark    {{\ensuremath{\Pc}}\xspace}

\def\bquark    {{\ensuremath{\Pb}}\xspace}

\def\pion   {{\ensuremath{\Ppi}}\xspace}

\def\pip    {{\ensuremath{\pion^+}}\xspace}
\def\pim    {{\ensuremath{\pion^-}}\xspace}

\def\rhomeson {{\ensuremath{\Prho}}\xspace}
\def\rhoz     {{\ensuremath{\rhomeson^0}}\xspace}

\def\kaon    {{\ensuremath{\PK}}\xspace}

\def\KorKbar {\kern \thebaroffset\optbar{\kern -\thebaroffset \PK}{}\xspace}

\def\Kp      {{\ensuremath{\kaon^+}}\xspace}
\def\Km      {{\ensuremath{\kaon^-}}\xspace}

\def\Kstarz  {{\ensuremath{\kaon^{*0}}}\xspace}

\newcommand{\omegaz}{\ensuremath{\Pomega}\xspace}

\def\Dbar    {{\ensuremath{\offsetoverline{\PD}}}\xspace}
\def\D       {{\ensuremath{\PD}}\xspace}

\def\DorDbar {\kern \thebaroffset\optbar{\kern -\thebaroffset \PD}\xspace}
\def\Dz      {{\ensuremath{\D^0}}\xspace}
\def\Dzb     {{\ensuremath{\Dbar{}^0}}\xspace}
\def\Dp      {{\ensuremath{\D^+}}\xspace}
\def\Dm      {{\ensuremath{\D^-}}\xspace}

\def\DpDm    {\ensuremath{\Dp {\kern -0.16em \Dm}}\xspace}

\def\Dstarp  {{\ensuremath{\D^{*+}}}\xspace}
\def\Dstarm  {{\ensuremath{\D^{*-}}}\xspace}

\def\Dsp     {{\ensuremath{\D^+_\squark}}\xspace}

\def\B       {{\ensuremath{\PB}}\xspace}

\def\BorBbar {\kern \thebaroffset\optbar{\kern -\thebaroffset \PB}\xspace}
\def\Bz      {{\ensuremath{\B^0}}\xspace}

\def\Bd      {{\ensuremath{\B^0}}\xspace}

\def\BdorBdbar {\kern \thebaroffset\optbar{\kern -\thebaroffset \Bd}\xspace}

\def\Bs      {{\ensuremath{\B^0_\squark}}\xspace}

\def\BsorBsbar {\kern \thebaroffset\optbar{\kern -\thebaroffset \Bs}\xspace}

\def\Y#1S{\ensuremath{\PUpsilon{(#1S)}}\xspace}

\def\proton      {{\ensuremath{\Pp}}\xspace}

\def\LorLbar     {\kern \thebaroffset\optbar{\kern -\thebaroffset \PLambda}\xspace}

\newcommand{\decay}[2]{\ensuremath{#1\!\to #2}\xspace} 

\def\to                 {\ensuremath{\rightarrow}\xspace}

\def\CP                {{\ensuremath{C\!P}}\xspace}

\def\AT#1     {\ensuremath{A_{\mathrm{T}}^{#1}}\xspace}           %

\def\C#1      {\ensuremath{\mathcal{C}_{#1}}\xspace}                       %
\def\Cp#1     {\ensuremath{\mathcal{C}_{#1}^{'}}\xspace}                    %
\def\Ceff#1   {\ensuremath{\mathcal{C}_{#1}^{\mathrm{(eff)}}}\xspace}        %
\def\Cpeff#1  {\ensuremath{\mathcal{C}_{#1}^{'\mathrm{(eff)}}}\xspace}       %
\def\Ope#1    {\ensuremath{\mathcal{O}_{#1}}\xspace}                       %
\def\Opep#1   {\ensuremath{\mathcal{O}_{#1}^{'}}\xspace}                    %

\newcommand{\aunit}[1]{\ensuremath{\text{\,#1}}}       
\newcommand{\tev}{\aunit{Te\kern -0.1em V}\xspace}
\newcommand{\gev}{\aunit{Ge\kern -0.1em V}\xspace}
\newcommand{\mev}{\aunit{Me\kern -0.1em V}\xspace}
\newcommand{\kev}{\aunit{ke\kern -0.1em V}\xspace}
\newcommand{\ev}{\aunit{e\kern -0.1em V}\xspace}
 
\newcommand{\mevc}{\ensuremath{\aunit{Me\kern -0.1em V\!/}c}\xspace}
\newcommand{\gevc}{\ensuremath{\aunit{Ge\kern -0.1em V\!/}c}\xspace}
\newcommand{\mevcc}{\ensuremath{\aunit{Me\kern -0.1em V\!/}c^2}\xspace}
\newcommand{\gevcc}{\ensuremath{\aunit{Ge\kern -0.1em V\!/}c^2}\xspace}

\def\fb   {\ensuremath{\aunit{fb}}\xspace}
\def\invfb   {\ensuremath{\fb^{-1}}\xspace}

\def\gsim{{~\raise.15em\hbox{$>$}\kern-.85em
          \lower.35em\hbox{$\sim$}~}\xspace}
\def\lsim{{~\raise.15em\hbox{$<$}\kern-.85em
          \lower.35em\hbox{$\sim$}~}\xspace}

\def\pt         {\ensuremath{p_{\mathrm{T}}}\xspace}

\def\tell1  {TELL1\xspace}
\def\ukl1   {UKL1\xspace}

\newcommand{\eg}{\mbox{\itshape e.g.}\xspace}

\usepackage{cite} %
\usepackage{mciteplus}
\usepackage{color}

\newcommand{\hhmm}{\ensuremath{h^+h^-\mu^+\mu^-}\xspace}

\newcommand{\Dhhmm}{\mbox{\ensuremath{\Dz\to h^+h^-\mu^+\mu^-}}\xspace}
\newcommand{\Dkkmm}{\mbox{\ensuremath{\Dz\to\Kp\Km\mu^+\mu^-}}\xspace}
\newcommand{\Dppmm}{\mbox{\ensuremath{\Dz\to\pip\pim\mu^+\mu^-}}\xspace}
\newcommand{\Dkpmm}{\mbox{\ensuremath{\Dz\to\Km\pip\mu^+\mu^-}}\xspace}

\newcommand{\mypaperversion}{}
\newcommand{\mydate}{June 9, 2022}

\newcommand{\mylhcbpapernumber}{LHCb-PAPER-2021-035}
\newcommand{\mycernpapernumber}{CERN-EP-2021-212}

\newcommand{\mmumu}{\ensuremath{m(\mup\mun)}\xspace}
\newcommand{\mhh}{\ensuremath{m(h^+h^-)}\xspace}
\newcommand{\mD}{\ensuremath{m(\hhmm)}\xspace}
\newcommand{\Acp}{\ensuremath{A_{\CP}}\xspace}
\newcommand{\hh}{\ensuremath{h^+h^-}\xspace}
\newcommand{\Araw}{\ensuremath{A_{\CP}^{\rm raw}}\xspace}
\newcommand{\Afb}{\ensuremath{A_{\mathrm{FB}}}\xspace}

\newcommand{\Si}{\ensuremath{\langle S_{\mathrm{i}}} \rangle \xspace}
\newcommand{\Sii}{\ensuremath{\langle S_{\mathrm{2}}} \rangle \xspace}
\newcommand{\Siii}{\ensuremath{\langle S_{\mathrm{3}}} \rangle \xspace}
\newcommand{\Siv}{\ensuremath{\langle S_{\mathrm{4}}} \rangle \xspace}
\newcommand{\Sv}{\ensuremath{\langle S_{\mathrm{5}}} \rangle \xspace}
\newcommand{\Svi}{\ensuremath{\langle S_{\mathrm{6}}} \rangle \xspace}
\newcommand{\Svii}{\ensuremath{\langle S_{\mathrm{7}}} \rangle \xspace}
\newcommand{\Sviii}{\ensuremath{\langle S_{\mathrm{8}}} \rangle \xspace}
\newcommand{\Six}{\ensuremath{\langle S_{\mathrm{9}}} \rangle \xspace}

\newcommand{\Ai}{\ensuremath{\langle A_{\mathrm{i}}} \rangle \xspace}
\newcommand{\Aii}{\ensuremath{\langle A_{\mathrm{2}}} \rangle \xspace}
\newcommand{\Aiii}{\ensuremath{\langle A_{\mathrm{3}}} \rangle \xspace}
\newcommand{\Aiv}{\ensuremath{\langle A_{\mathrm{4}}} \rangle \xspace}
\newcommand{\Av}{\ensuremath{\langle A_{\mathrm{5}}} \rangle \xspace}
\newcommand{\Avi}{\ensuremath{\langle A_{\mathrm{6}}} \rangle \xspace}
\newcommand{\Avii}{\ensuremath{\langle A_{\mathrm{7}}} \rangle \xspace}
\newcommand{\Aviii}{\ensuremath{\langle A_{\mathrm{8}}} \rangle \xspace}
\newcommand{\Aix}{\ensuremath{\langle A_{\mathrm{9}}} \rangle \xspace}

\newcommand{\Ii}{\ensuremath{\langle {I}_{\mathrm{i}}} \rangle \xspace}
\newcommand{\Iii}{\ensuremath{\langle I_{\mathrm{2}}} \rangle \xspace}

\newcommand{\Ibi}{\ensuremath{\langle \overline{{I}_{\mathrm{i}}}} \rangle \xspace}

\newcommand{\Aphi}{\ensuremath{A_{2\phi}}\xspace}

\usepackage{longtable} %
\usepackage{booktabs}
\usepackage{rotating}

\begin{document}

\renewcommand{\thefootnote}{\fnsymbol{footnote}}
\setcounter{footnote}{1}

%
%
%
%
%
%
%
%
%
%
%
%

%
%
%
\begin{titlepage}
\pagenumbering{roman}

\vspace*{-1.5cm}
\centerline{\large EUROPEAN ORGANIZATION FOR NUCLEAR RESEARCH (CERN)}
\vspace*{1.5cm}
\noindent
\begin{tabular*}{\linewidth}{lc@{\extracolsep{\fill}}r@{\extracolsep{0pt}}}
\ifthenelse{\boolean{pdflatex}}%
{\vspace*{-1.5cm}\mbox{\!\!\!\includegraphics[width=.14\textwidth]{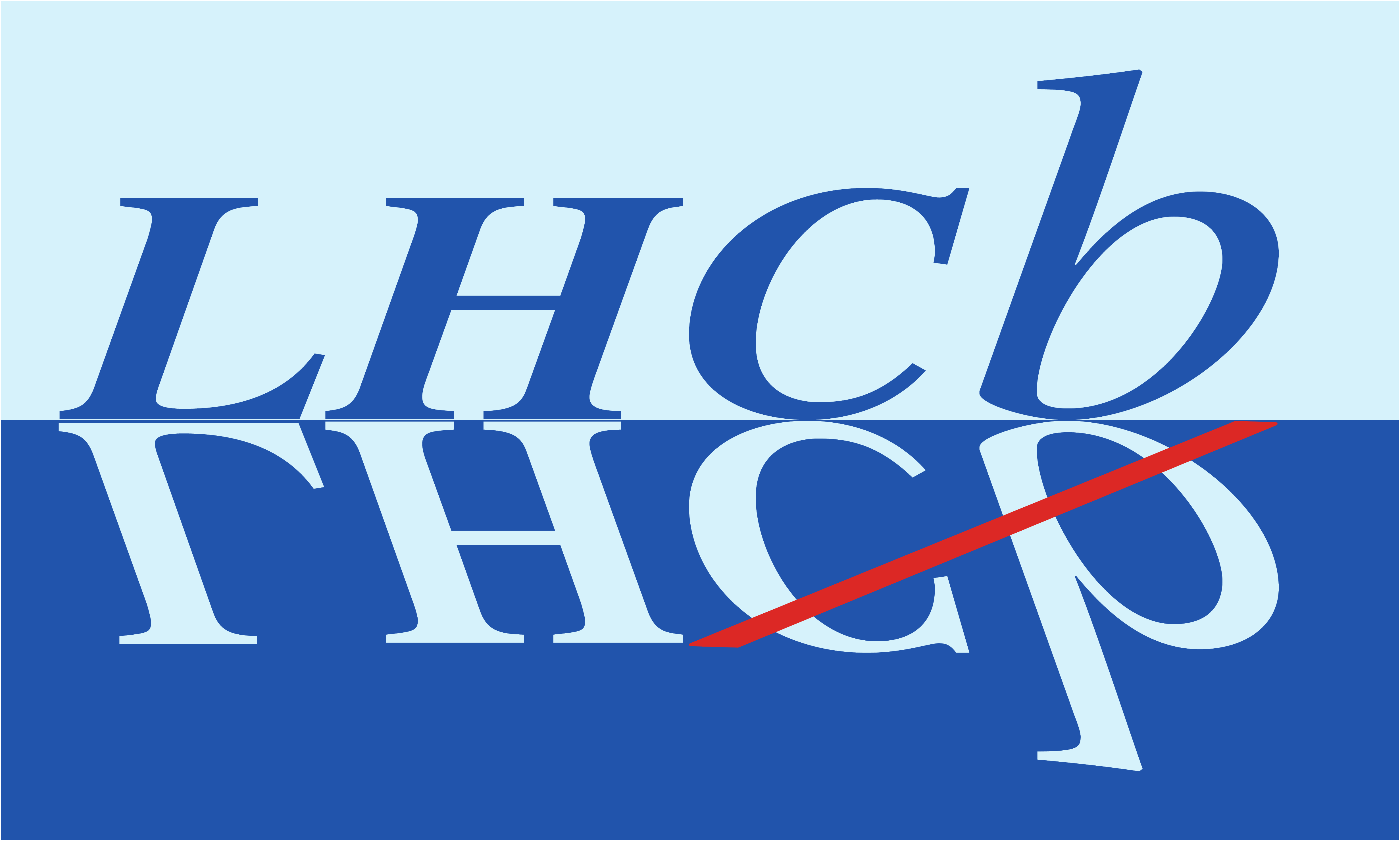}} & &}%
{\vspace*{-1.2cm}\mbox{\!\!\!\includegraphics[width=.12\textwidth]{figs/lhcb-logo.eps}} & &}%
\\
 & & \mycernpapernumber \\  %
 & & \mylhcbpapernumber \\  %
 & & \mydate \\ %
 & & \mypaperversion \\
\end{tabular*}

\vspace*{4.0cm}

{\normalfont\bfseries\boldmath\huge
\begin{center}
  \papertitle 
\end{center}
}

\vspace*{2.0cm}

\begin{center}
\paperauthors\footnote{Authors are listed at the end of this Letter.}
\end{center}

\vspace{\fill}

\begin{abstract}
%
\noindent
The first full angular analysis and an updated measurement of the decay-rate \CP asymmetry of the \Dppmm and \Dkkmm decays are reported. The analysis uses proton-proton collision data collected with the \lhcb detector at centre-of-mass energies of 7, 8, and 13\tev. The dataset corresponds to an integrated luminosity of 9\invfb. The full set of \CP-averaged angular observables and their \CP asymmetries are measured as a function of the dimuon invariant mass. The results are consistent with expectations from the standard model and with \CP symmetry.
 \end{abstract}

\vspace*{2.0cm}

\begin{center}
  Published in 
  Phys.~Rev.~Lett. 128, 221801 (2022)
\end{center}

\vspace{\fill}

{\footnotesize 
\centerline{\copyright~\papercopyright. \href{\paperlicenceurl}{\paperlicence}.}}
\vspace*{2mm}

\end{titlepage}

\newpage
\setcounter{page}{2}
\mbox{~}
%
%
%
%
%
%
%
%
%
%
%
%
%
%
%
%
%
%
%
%

 %
%

\renewcommand{\thefootnote}{\arabic{footnote}}
\setcounter{footnote}{0}

\cleardoublepage

\pagestyle{plain} %
\setcounter{page}{1}
\pagenumbering{arabic}

%
%
%

%
%
%
%
Rare charm decays with two oppositely charged leptons ($\ell^+\ell^-$) in the final state may proceed via the quark flavor-changing neutral-current (FCNC) process $\cquark\to\uquark\ell^+\ell^-$ and, as such, be sensitive to contributions from physics beyond the standard model (SM). They represent a unique probe to beyond-SM couplings to up-type quarks, which is complementary to recent studies of beauty quark $\bquark\to\squark\ell^+\ell^-$ transitions, where a coherent pattern of deviations from the SM is emerging (\eg, see Refs.~\cite{LHCb-PAPER-2021-014,LHCB-PAPER-2021-004,LHCB-PAPER-2020-002}). The loop-induced SM processes are more suppressed in charm than in the \bquark-quark system due to the Glashow-Iliopoulos-Maiani mechanism~\cite{GIM}. The {\em short-distance} contributions to the inclusive $\D\to X\mup\mun$ branching fraction, where $\D$ denotes a neutral or charged $\D$ meson and $X$ represents one or more hadrons, are predicted to be of $\mathcal{O}(10^{-9})$~\cite{PaulBigi:2011}. Sensitivity to FCNC processes via the measurement of branching fractions is limited due to the dominance of tree-level amplitudes involving intermediate resonances that subsequently decay into $\ell^+\ell^-$. These so-called {\em long-distance} contributions increase the SM branching fractions up to $\mathcal{O}(10^{-6})$~\cite{Fajfer:2007,PaulBigi:2011,Cappiello,deBoer:2018}. Studies of angular distributions and charge-parity (\CP) asymmetries in the vicinity of intermediate resonances offer a access to  observables with negligible theoretical uncertainties. These observables are sensitive to beyond-SM physics through the interference between long- and short-distance amplitudes. The values of these observables are negligibly small in the SM, but can reach the percent level in scenarios beyond the SM~\cite{Fajfer:2005ke,Bigi:2012,Paul:2012ab,Fajfer:2012nr,Cappiello,Fajfer:2015mia,deBoer:2015boa,deBoer:2018,Gisbert:2020vjx,Bause:2019vpr,Bause:2020obd,Bharucha:2020eup}.

The \lhcb Collaboration has previously reported the first observation of \Dhhmm decays, where $h$ is either a pion or a kaon~\cite{LHCb-PAPER-2017-019}. Charge-conjugate decays are implied throughout, unless stated otherwise. The measured branching fractions are in agreement with SM predictions~\cite{Cappiello,deBoer:2018}. Selected angular and \CP asymmetries were also measured, with results in agreement with the SM and with \CP symmetry~\cite{LHCb-PAPER-2018-020}. However, a complete angular analysis of a rare charm decay is yet to be performed.

This Letter presents the first measurement of the full set of \CP-averaged angular observables and  \CP asymmetries in \Dhhmm decays, together with an updated measurement of the \CP asymmetry of the total decay rate, defined as 
\begin{align} 
\Acp\equiv \frac{\Gamma(\Dz \to h^+ h^- \mumu) - \Gamma(\Dzb \to h^+ h^- \mumu)}{\Gamma(\Dz \to h^+ h^- \mumu) + \Gamma(\Dzb \to h^+ h^- \mumu)}\,,
\end{align}
where $\Gamma$ indicates the total decay rate. The measurement uses proton-proton (\proton\proton) collision data corresponding to an integrated luminosity of 9\invfb, collected by the \lhcb experiment at center-of-mass energies of 7 and 8\tev (run~1) and of 13\tev (run~2). The analysis is an extension of that reported in  Ref.~\cite{LHCb-PAPER-2018-020}. It uses approximately three times as many signal decays and results for previously measured observables are superseded.

Semileptonic \Dhhmm decays are described by five independent kinematic variables: the squared invariant masses of the dimuon and dihadron systems, \mbox{$q^2\equiv m^2(\mumu)$} and \mbox{$p^2\equiv m^2(\hh)$}, and three decay angles $\theta_\mu, \theta_h, \phi$, (see Fig.~S1 in the Supplemental Material~\cite{supplemental}). Here, $\theta_\mu$ is the angle between the $\mu^+$ direction and the direction opposite to that of the \Dz meson in the dimuon rest frame; $\theta_h$ is the angle between the $h^+$ direction and the direction opposite to that of the \Dz meson in the dihadron rest frame; and $\phi$ is the angle between the two planes formed by the dimuon and the dihadron systems in the rest frame of the \Dz meson~\cite{deBoer:2018,LHCb-PAPER-2018-020,supplemental}.
 In contrast to Ref.~\cite{LHCb-PAPER-2018-020}, the same definition of the angles is kept for \Dz and \Dzb mesons.
Following Ref.~\cite{deBoer:2018} and defining $\Vec{\Omega}\equiv (\cos\theta_\mu,\cos\theta_h,\phi)$, the differential decay rate is expressed as the sum of nine angular coefficients $I_{1-9}$ that depend on $q^2, p^2$, and $\cos \theta_h$, multiplied by the terms $c_1=1$, $c_2=\cos 2\theta_\mu$, $c_3=\sin^2\theta_\mu\cos 2\phi$, $c_4=\sin 2\theta_\mu \cos \phi$, $c_5=\sin\theta_\mu\cos\phi$, $c_6=\cos\theta_\mu$, $c_7=\sin\theta_\mu\sin\phi$, $c_8=\sin 2\theta_\mu\sin\phi$, and $c_9=\sin^2\theta_\mu\sin2\phi$, as
\begin{align} \label{supp:eq:diffBF_appendix}
\frac{d^5\Gamma }{dq^2\,dp^2\,d\Vec{\Omega}} =&\frac{1}{ 2  \pi} \sum_{i=1}^{9} c_i\, {I}_i \,.
\end{align}
Piecewise integration ranges in $\phi$ and $\cos\theta_{\mu}$ can be defined such that the coefficients ${I}_{2-9}$ are expressed as angular asymmetries. For example, the coefficient ${I}_{2}$ is obtained as
\begin{align} \label{eq:I2ToI4} \begin{split}
{I}_2  & = \int_{-\pi}^\pi  d\phi \, \left[ \int_{-1}^{-0.5} d\cos\theta_{\mu} + \int_{0.5}^{1} d\cos\theta_{\mu} - \int_{-0.5}^{0.5}  d\cos\theta_{\mu}   \right]  \, \frac{d^5\Gamma }{dq^2\,dp^2\,d\Vec{\Omega}}\,. \\
\end{split}  \end{align} 
Corresponding integration ranges to obtain $I_{3-9}$ are reported in Refs.~\cite{deBoer:2018,supplemental}. The coefficients $I_{2,3,4,7}$ are even under \CP transformations, while $I_{5,6,8,9}$ are \CP odd. Since the term $c_1$ has no dependence on the decay angles, $I_1$ provides only a normalization factor and is not considered in this Letter. 

This analysis measures the normalized observables $\langle I_{2-9}\rangle$ defined as 
\begin{align} \label{eq:averagedCoefficients} \begin{split}
 \langle I_{2,3,6,9} \rangle  & = \frac{1}{{\Gamma}} \int_{q^2_{\min}}^{q_{\max}^2} dq^2 \int_{p_{\min}^2}^{p_{\max}^2} d p^2 \int_{-1}^{+1} d \cos \theta_h \, {I}_{2,3,6,9} \, , \\
 \langle {I}_{4,5,7,8} \rangle  & = \frac{1}{{\Gamma}} \int_{q^2_{\min}}^{q_{\max}^2} dq^2  \int_{p_{\min}^2}^{p_{\max}^2} d p^2  \left[ \int_{0}^{+1} d \cos \theta_h  -  \int_{-1}^{0} d \cos \theta_h  \right] \,{I}_{4,5,7,8} \,,\\
\end{split}  \end{align} 
where $\Gamma$ is the decay rate in the considered region of dimuon mass. The integration boundaries $q^2_{\min}$, $q^2_{\max}$, and $p^2_{\max}$ depend on the dimuon-mass region, where $p^2_{min} = 4 m_h^2$ and $m_{h}$ denotes the hadron mass. The integration in $\cos \theta_h$ is defined to optimize the sensitivity to beyond-SM effects by integrating out contributions from the dominant P-wave resonances in the dihadron system, which further decay into $h^+h^-$~\cite{deBoer:2018}. Experimentally, the observables are determined by measuring the decay-rate asymmetries of the data split by angular \textit{tags} defined according to the piecewise integration of the decay rate. As an example, from Eqs.~\eqref{eq:I2ToI4} and \eqref{eq:averagedCoefficients}, $\Iii$ is measured as
\begin{align} 
\Iii = \frac{1}{\Gamma}\bigl[\Gamma(|\cos \theta_\mu| >0.5) - \Gamma(|\cos \theta_\mu | <0.5)\bigr]\,.
\end{align} 
The observables $\Ii$, measured separately for \Dz and \Dzb mesons, are labeled as $\Ii$ and $\Ibi$, respectively. Their \CP average $\Si$ and asymmetry $\Ai$ are defined as \mbox{$\Si= \frac{1}{2} \left[ \Ii +(-) \Ibi \right]$} and \mbox{$\Ai = \frac{1}{2} \left[ \Ii  -(+) \Ibi \right]$} for the \CP-even (\CP-odd) coefficients $\langle I_{2,3,4,7} \rangle$ ($\langle I_{5,6,8,9} \rangle$). The previously measured forward-backward asymmetry \Afb and triple-product asymmetry \Aphi~\cite{LHCb-PAPER-2018-020} are related to $\Svi$ and $\Aix$, respectively. If only SM amplitudes contribute to the decay processes, the observables $\langle S_{5,6,7} \rangle$ are predicted to vanish and constitute SM null tests together with the \CP asymmetries $\langle A_{2-9} \rangle$, which are expected to be below the current experimental sensitivity~\cite{deBoer:2018}. No predictions are available for the observables $\langle S_{2,3,4,8,9} \rangle$.

The analysis is performed using \Dz mesons that originate from decays of \Dstarp mesons directly produced in the primary \proton\proton collision. The charge of the pion in the decay chain $\Dstarp \to \Dz \pi^+$ is used to infer the flavor of the \Dz meson at its production. All observables are measured integrated in the full $\mmumu$ range and in $\mmumu$ regions defined according to the presence of the known intermediate resonances~\cite{LHCb-PAPER-2018-020}. For \Dppmm decays the regions are (low-mass) below $525\mevcc$, ($\eta$) $525$--$565\mevcc$, ($\rhoz/\omegaz$ low) $565$--$780\mevcc$, ($\rhoz/\omegaz$ high) $780$--$950\mevcc$, ($\phi$ low) $950$--$1020\mevcc$, ($\phi$ high) $1020$--$1100\mevcc$, and (high mass) above $1100\mevcc$. For \Dkkmm decays, three regions are considered: (low mass) below $525\mevcc$, ($\eta$) $525$--$565\mevcc$, and only one region that combines the low and high $\rhoz/\omegaz$ region due to limited signal yields. The asymmetries are determined only in \mmumu regions where a significant signal yield was previously observed~\cite{LHCb-PAPER-2017-019}. No measurement is performed in the $\eta$ region of both channels and in the high-mass region of \Dppmm. The integrated measurement includes candidates from all \mmumu intervals. For each \mmumu region, the kinematically allowed $m(h^+h^-)$ range up to a maximum of $1200\mevcc$, is considered. To avoid potential experimenter bias on the measured quantities, the observables were shifted by an unknown value during the development of the analysis and examined only after the analysis procedure was finalized.

The \lhcb detector is a single-arm forward spectrometer described in detail in Ref.~\cite{LHCb-DP-2008-001}. It provides high-precision tracking and good particle identification over a large range in momentum~\cite{LHCb-DP-2014-002}. 
Simulation~\cite{Sjostrand:2006za,Sjostrand:2007gs,LHCb-PROC-2010-056,Lange:2001uf,davidson2015photos,Allison:2006ve, Agostinelli:2002hh,LHCb-PROC-2011-006} is used to optimize the selection and to estimate variations of the reconstruction and selection efficiency across the decay phase space. Corrections to account for mismodeling of the charged-particle multiplicity of the events and of the particle-identification performance are applied using control channels in the  data~\cite{LHCb-PUB-2016-021,LHCb-DP-2018-001}.

Events are selected online by a trigger that consists of a hardware stage, based on information from the muon systems, followed by a software stage, based on charged tracks that are displaced from any primary $pp$-interaction vertex (PV). A subsequent software trigger exploits a full event reconstruction~\cite{LHCb-DP-2012-004} to select \Dhhmm candidates. Online selection requirements that have changed over data-taking periods are equalized in the off-line selection, which follows closely that of Ref.~\cite{LHCb-PAPER-2018-020}. 

Candidate \Dz mesons are constructed off-line by combining four charged tracks, each having momentum larger than $3000\mevc$ and the momentum component transverse to the beam direction $\pt>300\mevc$, that form a good-quality secondary vertex (SV) significantly displaced from any PV in the event. Two oppositely charged particles are required to be identified as muons and two as either pions or kaons. The \Dz candidates are required to have invariant mass in the range \mbox{$1820 < \mD < 1940$\mevcc} and to be consistent with originating from the associated PV, defined as the PV with respect to which the \Dz candidate has the lowest impact-parameter significance. The \Dz momentum is required to be aligned with the vector connecting the PV and the SV. The mass of the dihadron system is required to be less than $1200\mevcc$. The \Dz candidates are combined with low-momentum charged pions having $\pt>120\mevc$, denoted as soft pions in the following, to form \Dstarp candidates. The \Dstarp decay vertex is required to coincide with the position of the associated PV. The difference between the masses of the \Dstarp and \Dz candidates, $\Delta m$, must be within \mbox{$|\Delta m-145.4 \mevcc|<0.6\mevcc$}, corresponding to approximately $\pm2$ standard deviations in $\Delta m$ resolution around the known value~\cite{PDG2020}. 

To suppress the combinatorial background formed with randomly associated tracks that accidentally fulfil the selection requirements, a boosted decision tree (BDT) algorithm~\cite{Breiman,Roe} with gradient boosting~\cite{TMVA} is employed. Simulated decays and data candidates from the sideband region $\mD>1890\mevcc$ are used as signal and background proxies, respectively. To have an unbiased estimate of the BDT performance, a cross validation is performed. The training samples are randomly split into two halves and the BDT classifier is applied to the subsample that has not been used in the training. Separate classifiers are trained for \Dppmm and \Dkkmm decays and for run~1 and run~2 data samples to account for differences in decay kinematics and data-taking conditions, respectively. The variables used in the training are momentum and $\pt$ of the soft pion, the largest distance of closest approach of the \Dz decay-product trajectories, the angle between the \Dz momentum and the vector connecting the PV and the SV, the fit quality of the SV and its spatial separation from the PV. Purely hadronic decays of the form $\Dz \to h^+ h^- \pip \pim$ with two pions wrongly identified as muons are further reduced by requirements on muon identification~\cite{Archilli:2013npa,Aaij:2018vrk}. 
The optimal working points of the BDT output selection and muon-identification thresholds are determined simultaneously by maximizing the quantity $\mathcal{S}/\sqrt{\mathcal{S}+\mathcal{B}}$, where $\mathcal{S}$ and $\mathcal{B}$ are the signal and background yields, respectively, determined from the data in the signal region defined as \mbox{$1840 < \mD < 1890 \mevcc$}. In the approximately $0.5\%$ of events where multiple \Dppmm candidates are reconstructed after the full selection, only one is kept at random. No multiple-candidate events are found for \Dkkmm decays.

\begin{figure}[t]
\centering
\includegraphics[width=.5\textwidth]{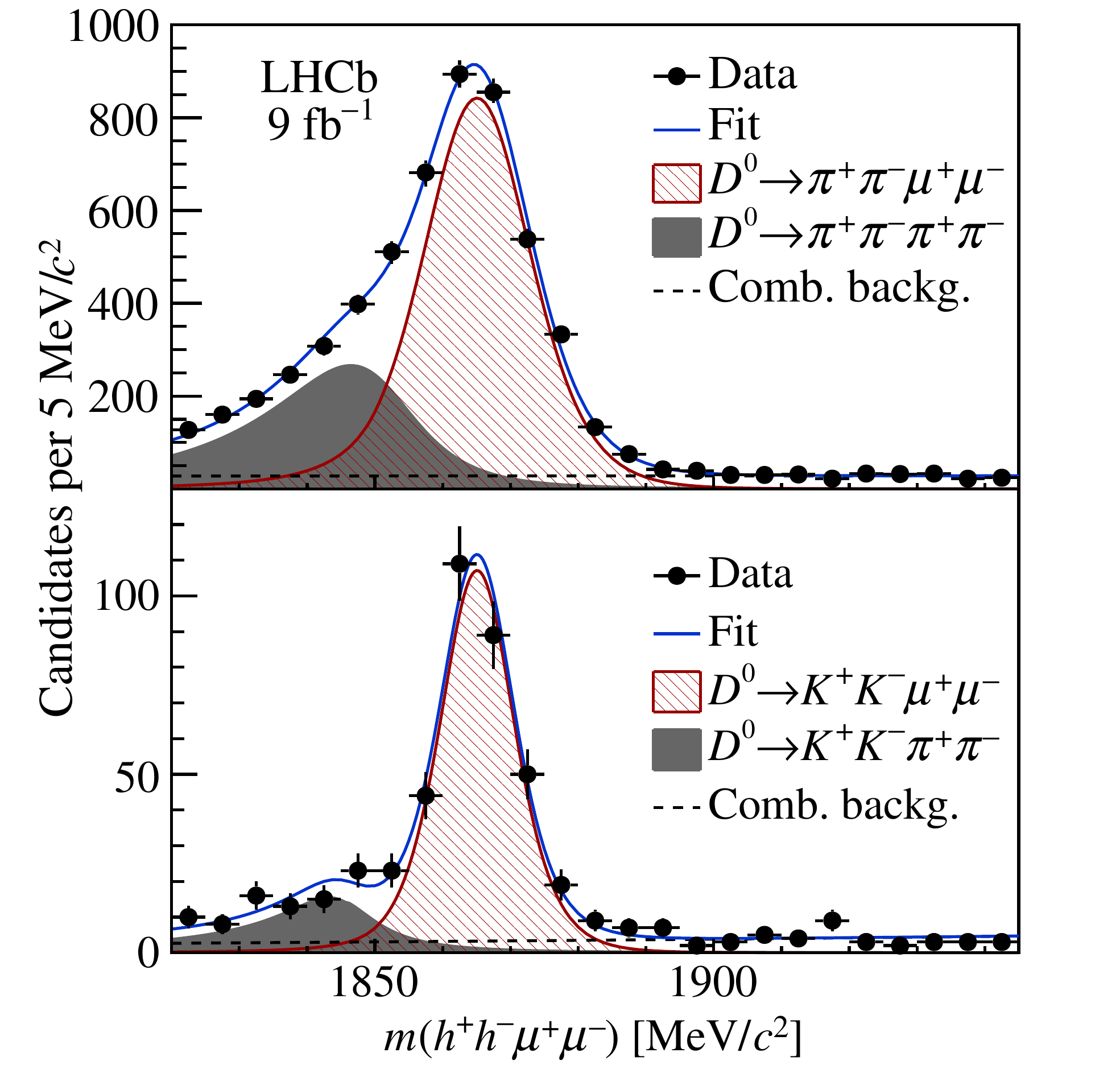}\\
\caption{Mass distribution of (top) \Dppmm and (bottom) \Dkkmm candidates with fit projections overlaid.}\label{fig:DataSignalFits}
\end{figure} 

The \mD distributions for selected candidates are shown in Fig.~\ref{fig:DataSignalFits}. Unbinned maximum-likelihood fits to these distributions yield \mbox{$3579 \pm 71$} \Dppmm and \mbox{$318 \pm 19$} \Dkkmm signal decays. The signal probability density function (PDF) is described by a Hypatia distribution~\cite{Santos:2013gra} with parameters fixed from simulation, apart from two factors scaling the width and mean of the distribution to account for data-simulation differences. Misidentified hadronic decays are described by a Johnson $S_U$ distribution~\cite{johnson} with parameters fixed from a fit to high-yield data samples of $\Dz\to h^+h^-\pip\pim$ decays with muon-mass hypothesis assigned to two pions and muon-identification criteria applied only to one of them. The combinatorial background is described by an exponential function with shape fixed from a fit to the candidates satisfying \mbox{$|\Delta m-145.4 \mevcc|>2\mevcc$}, \mbox{$\Delta m< 170 \mevcc$}, and $1880 < \mD < 1945\mevcc$. 

The \lhcb detector geometry, signal reconstruction, and selection requirements result in nonuniform  efficiency across the five-dimensional phase space of the decays defined by $p^2$, $q^2$, $\theta_\mu$, $\theta_h$, and $\phi$. The efficiency variations are corrected using a method developed in Refs.~\cite{Viaud2016,LHCb-PAPER-2018-020}. A BDT classifier with gradient boosting~\cite{Breiman,Roe,TMVA} is used to identify differences in the decay kinematics before and after reconstruction and selection. The BDT classifier is trained on simulation using the five-dimensional phase-space variables as input. From the classifier output, per-event candidate weights are derived that correspond to the inverse efficiency. To account for the different detector conditions, separate classifiers are trained for run~1 and run~2 data samples. As a consequence of the weighting, the effective statistical power of the \Dppmm (\Dkkmm) data sample is reduced by approximately $10\%$ ($6\%$).

To determine the \CP asymmetry \Acp, the raw asymmetry in \Dz- and \Dzb-signal yields, \Araw, is corrected for $\mathcal{O}(1\%)$ nuisance asymmetries: differences in the production cross section of \Dstarp and \Dstarm mesons, $A_P$, and in the detection efficiencies of positively and negatively charged soft pions, $A_D$. The raw asymmetry for decays to the final state $f$ is approximated as $\Araw(f) \approx \Acp(f) + A_P + A_D$. A high-yield control sample of \mbox{$\Dstarp\to\Dz(\rightarrow K^+K^-)\pip$} decays is used to  determine the combined  nuisance asymmetry as \mbox{$A_P + A_D \approx \Araw (\Kp\Km) - \Acp(\Kp\Km)$}, using \mbox{$\Acp(\Kp\Km)=(-0.06\pm0.18)\%$} from the independent measurement of Ref.~\cite{LHCb-PAPER-2014-013}. In this procedure, the two-dimensional distribution of momentum and pseudorapidity of \Dstarp candidates in the control samples is equalized to that of the signal decays to account for differences in decay kinematics. Since the angular observables are measured independently for \Dz and \Dzb mesons, no correction for nuisance asymmetries is needed. 

\begin{figure*}[t]
\centering
\includegraphics[width=.5\textwidth]{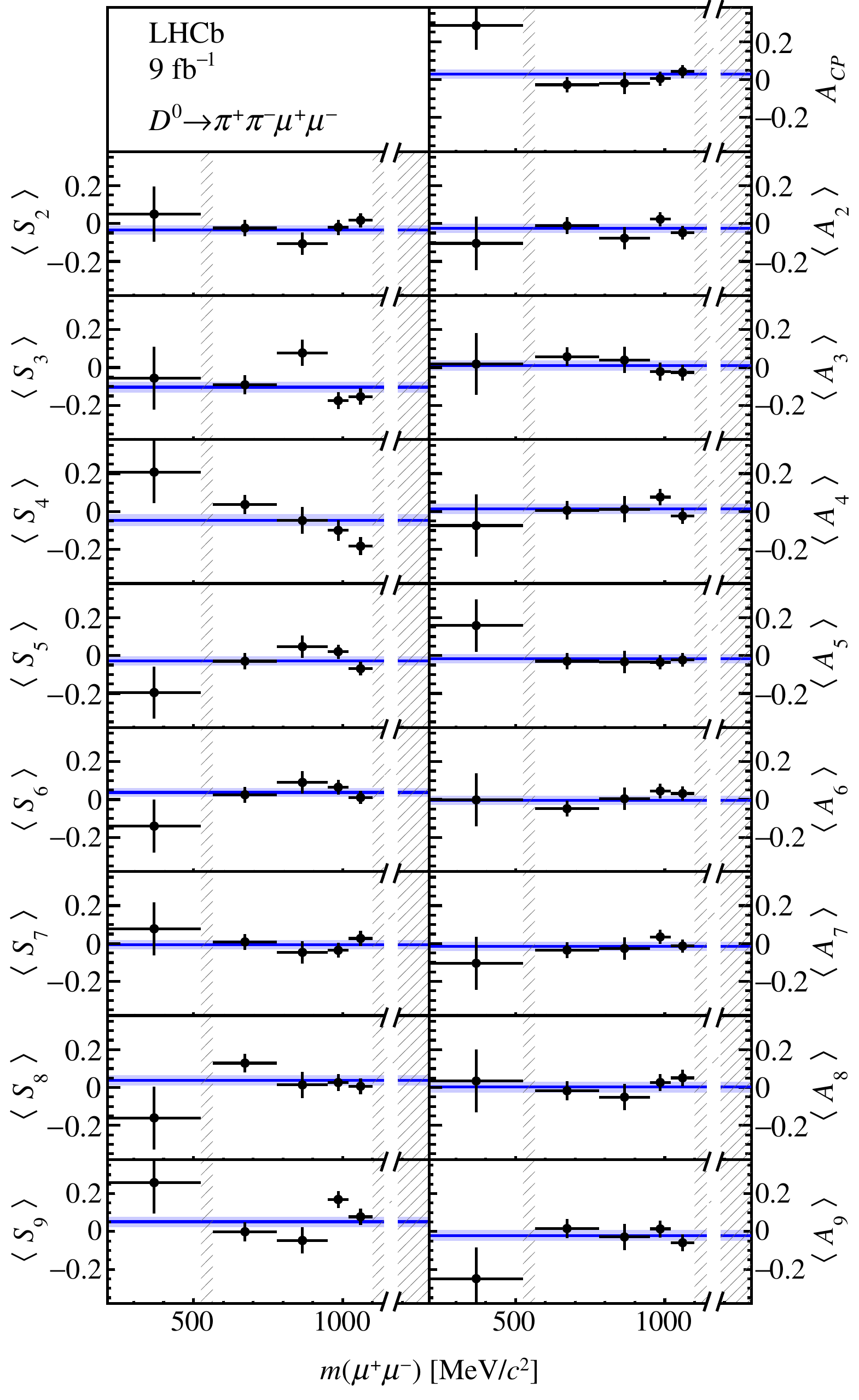}\hfil
\includegraphics[width=.5\textwidth]{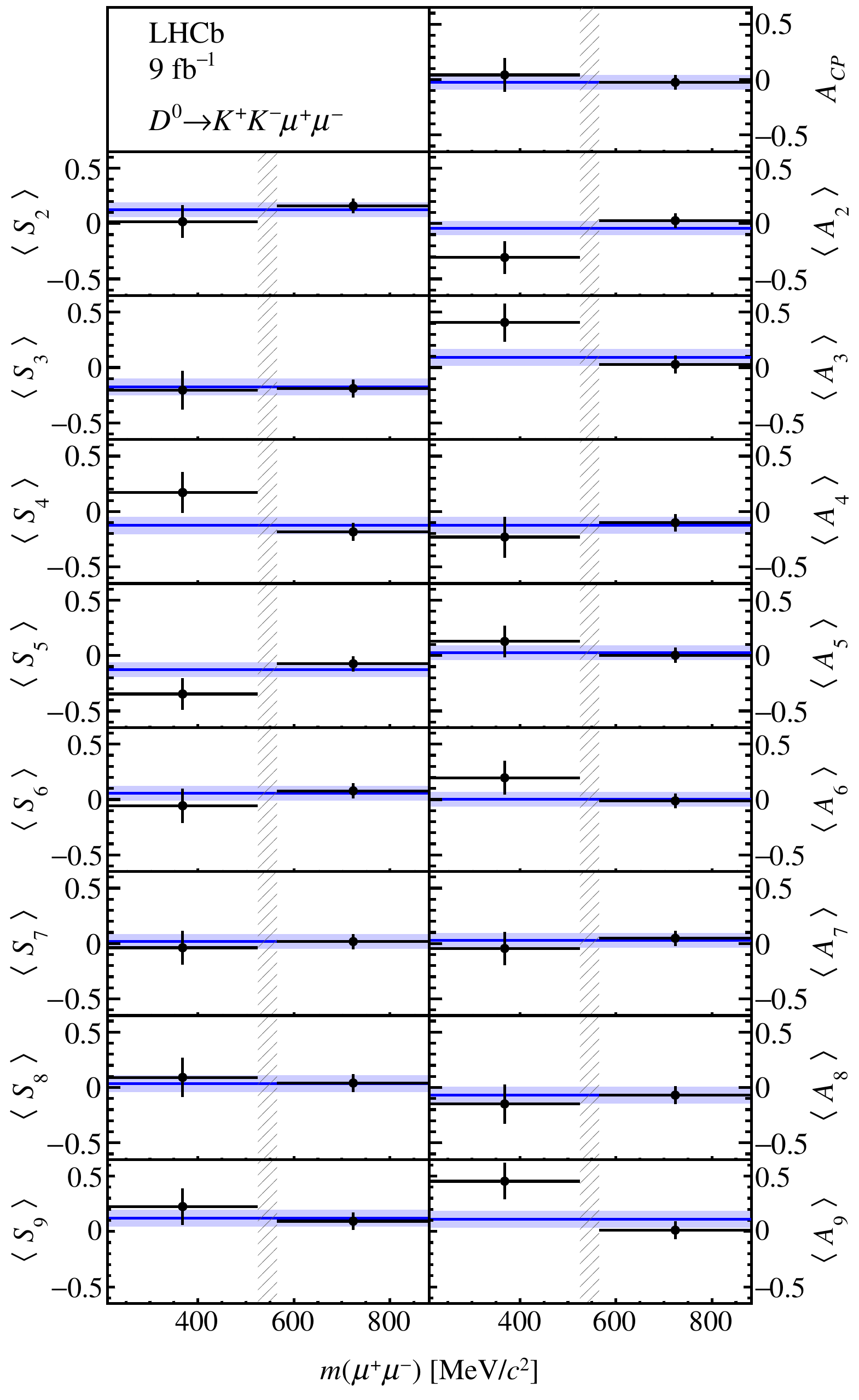}\\
\caption{Measured observables for (left) \Dppmm and (right) \Dkkmm decays in \mmumu regions.
No measurement is performed in the regions indicated by the vertical gray bands. The horizontal bands correspond to the measurements integrated in the dimuon mass, including candidates from all \mmumu ranges. The high-mass region of \Dppmm extends to $1590.5\mevcc$ and has been truncated on the plots for a clearer visualization of the other regions. \label{fig:Results}}
\end{figure*}

Each angular observable (or \Araw) is determined, independently in each dimuon-mass region, through simultaneous unbinned maximum-likelihood fits to the efficiency-corrected \mD distributions of candidates split according to the angular tag (or \Dz-meson flavor). The fits use the same model as described earlier, but with PDFs determined independently in each dimuon-mass region. The same PDFs are assumed when fitting the subsamples split by the tags, except for the measurement of $\Iii$, where the mass shape of misidentified hadronic decays depends on the angular tag. The yield and angular observable (or \Araw) of the three fit components (signal, misidentified, and combinatorial background) are the only floating parameters. The results for $\langle S_i \rangle$, $\langle A_i \rangle$, and \Acp, including both statistical and systematic uncertainties added in quadrature, are reported in Fig.~\ref{fig:Results}. In general, the null-test observables show agreement with the SM predictions. A tabulated version is given in the Supplemental Material~\cite{supplemental} together with the correlations between the observables, estimated using a bootstrapping technique~\cite{efron:1979}.

Systematic uncertainties are typically between 10\% and 50\% of the statistical uncertainty, depending on the observable and the dimuon-mass region. These arise from the following sources: the model used in the mass fits; neglected background from partially reconstructed \Dsp mesons, from \Dstarp candidates made of \Dz mesons combined with unrelated soft pions, and from \Dstarp candidates originating from decays of $\bquark$-flavored hadrons; uncertainties in the estimation of the efficiency correction; the accuracy of the correction for nuisance charge asymmetries; the finite resolution of the angular variables. The leading systematic uncertainties are those related to the efficiency correction. These include residual biases on the observables due to efficiency variations that are not fully accounted for by the correction, the uncertainty coming from the limited size of the simulation sample, and effects due to potential residual differences between the data and simulation. The systematic uncertainties due to the efficiency correction are evaluated by repeating the analysis on either fully simulated decays or on simplified simulations that mimic the presence of data-simulation differences. The generation of the simplified simulations reproduces the \mmumu and \mhh distributions observed in the data and is performed exploiting multithreaded architectures using the Hydra library~\cite{hydra,Alves:2017zzi}.

The analysis procedure is validated using the more abundant $\Dkpmm$ decay in the dimuon-mass range \mbox{$675 < \mmumu < 875 \mevcc$}, where the contribution from the \mbox{$\rhoz/\omega \to \mumu$} decay is dominant. The decay is not sensitive to FCNC processes and is dominated by the SM tree-level amplitude. The analysis measures the angular observables serving as SM null test ($\langle A_{2-9} \rangle$ and $\langle S_{5-7} \rangle$) to be consistent with zero within approximately $1\%$. As a further cross-check, the analysis is repeated on disjoint subsamples of the data selected according to criteria such as the magnetic-field orientation, which is reversed periodically during data taking; the number of PVs in the event; the transverse momentum of the \Dstarp and soft-pion candidates; and the minimum distance of the \Dz meson to the PV. The resulting variations of the measured observables are as expected according to statistical variations.

In summary, a measurement of the full set of \CP-averaged angular observables and their \CP asymmetries in \Dppmm and \Dkkmm decays is reported, together with an updated measurement of \Acp. The analysis uses \proton\proton collision data collected with the \lhcb detector at center-of-mass energies of 7, 8, and 13\tev, corresponding to an integrated luminosity of 9\invfb. This is the first full angular analysis of a rare charm decay ever performed. The measured null-test observables $\Acp$, $\langle S_{5-7}\rangle$ and $\langle A_{2-9} \rangle$ are in agreement with the SM null hypothesis with overall $p$ values of 79\% (0.8\%) for \Dppmm (\Dkkmm) decays, corresponding to 0.3 (2.7) Gaussian standard deviations. These measurements will help constraining the parameters space of physics models extending the SM. 

 
%
%
%
%
\section*{Acknowledgements}
\noindent We express our gratitude to our colleagues in the CERN
accelerator departments for the excellent performance of the LHC. We
thank the technical and administrative staff at the LHCb
institutes.
We acknowledge support from CERN and from the national agencies:
CAPES, CNPq, FAPERJ and FINEP (Brazil); 
MOST and NSFC (China); 
CNRS/IN2P3 (France); 
BMBF, DFG and MPG (Germany); 
INFN (Italy); 
NWO (Netherlands); 
MNiSW and NCN (Poland); 
MEN/IFA (Romania); 
MSHE (Russia); 
MICINN (Spain); 
SNSF and SER (Switzerland); 
NASU (Ukraine); 
STFC (United Kingdom); 
DOE NP and NSF (USA).
We acknowledge the computing resources that are provided by CERN, IN2P3
(France), KIT and DESY (Germany), INFN (Italy), SURF (Netherlands),
PIC (Spain), GridPP (United Kingdom), RRCKI and Yandex
LLC (Russia), CSCS (Switzerland), IFIN-HH (Romania), CBPF (Brazil),
PL-GRID (Poland) and NERSC (USA).
We are indebted to the communities behind the multiple open-source
software packages on which we depend.
Individual groups or members have received support from
ARC and ARDC (Australia);
AvH Foundation (Germany);
EPLANET, Marie Sk\l{}odowska-Curie Actions and ERC (European Union);
A*MIDEX, ANR, IPhU and Labex P2IO, and R\'{e}gion Auvergne-Rh\^{o}ne-Alpes (France);
Key Research Program of Frontier Sciences of CAS, CAS PIFI, CAS CCEPP, 
Fundamental Research Funds for the Central Universities, 
and Sci. \& Tech. Program of Guangzhou (China);
RFBR, RSF and Yandex LLC (Russia);
GVA, XuntaGal and GENCAT (Spain);
the Leverhulme Trust, the Royal Society
 and UKRI (United Kingdom).

 
\addcontentsline{toc}{section}{References}
\setboolean{inbibliography}{true}
\bibliographystyle{LHCb}
\ifx\mcitethebibliography\mciteundefinedmacro
\PackageError{LHCb.bst}{mciteplus.sty has not been loaded}
{This bibstyle requires the use of the mciteplus package.}\fi
\providecommand{\href}[2]{#2}

\setboolean{inbibliography}{false}
%
\clearpage
\renewcommand{\thefigure}{S\arabic{figure}}
\renewcommand{\thetable}{S\arabic{table}}
\renewcommand{\theequation}{S\arabic{equation}}

\setcounter{equation}{0}
\setcounter{figure}{0}
\setcounter{table}{0}
\setcounter{page}{1}

\section*{Supplemental material for the Letter ``Angular analysis of \Dppmm and \Dkkmm decays and search for \CP violation''}

A summary of the formalism used to describe the angular distribution of the \Dppmm and \Dkkmm decays, with the definition of all measured angular observables is reported below, followed by the full set of results in tabular form.

\begin{figure}[tb]
\centering
\includegraphics[width=0.7\linewidth]{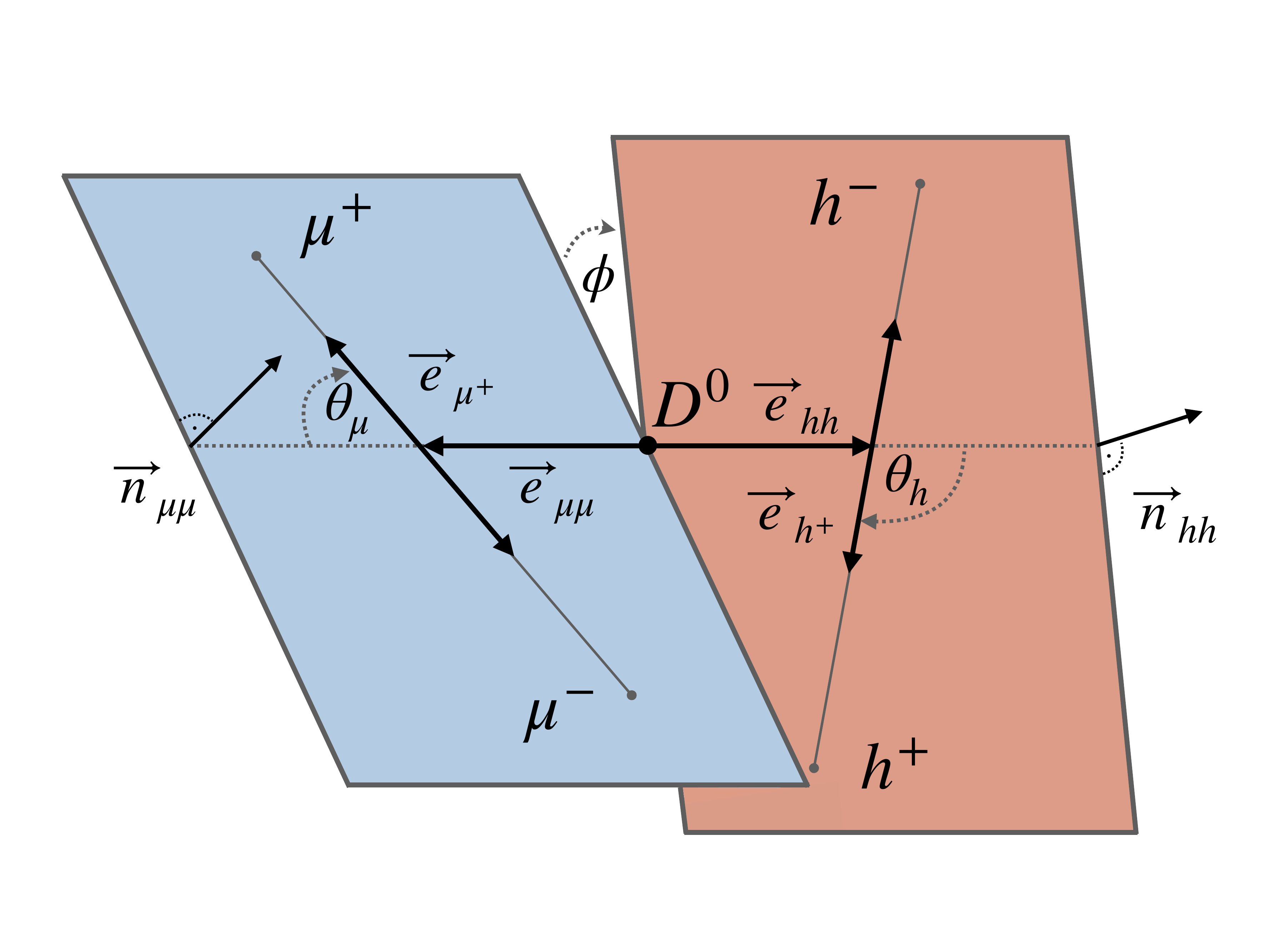}\hfil
\caption{Decay topology of \Dhhmm decays illustrating the definition of the decay angles $\theta_{\mu}$, $\theta_{h}$ and $\phi$.}\label{fig:sigTopology}
\end{figure}

The angular distribution of $\Dz \to h^+ h^- \mup \mun$ ($h=\pi,K$) decays can be written as~\cite{deBoer:2018}
\begin{align} \label{supp:eq:diffBF_appendix}
\frac{d^5\Gamma }{dq^2\,dp^2\,d\Vec{\Omega}} =&\frac{1}{ 2  \pi} \left[ \sum_{i=1}^{9} c_i(\theta_\mu,\phi) {I}_i (q^2,p^2,\cos \theta_{h}) \right]\,,
\end{align}
with the angular basis, $c_i$, defined as
\begin{align}\label{eq:ci}
c_1 & =1\,, \;  c_2=\cos 2\theta_\mu\,, \; c_3=\sin^2\theta_\mu\cos 2\phi\,, \; c_4=\sin 2\theta_\mu \cos \phi\,, \; c_5=\sin\theta_\mu\cos\phi\,, \nonumber \\ c_6& =\cos\theta_\mu\,, \; c_7=\sin\theta_\mu\sin\phi\,, \; c_8=\sin 2\theta_\mu\sin\phi\,, \; c_9=\sin^2\theta_\mu\sin2\phi \, . 
\end{align}
The variable $\cos \theta_{\mu}$ ($\cos \theta_{h}$) is the cosine of the angle between the momentum of the positive muon (hadron) in the rest frame of the dimuon (dihadron) system with respect to the dimuon (dihadron) flight direction as seen from the rest frame of the $\Dz$ candidate (see Fig.~\ref{fig:sigTopology}):
\begin{equation}
\begin{aligned}   
\cos{\theta_{\mu}} &= \vec{e}_{\mu\mu} \cdot \vec{e}_{\mu^{+}}, \\
\cos{\theta_{h}} &= \vec{e}_{hh} \cdot \vec{e}_{h^{+}}. 
\end{aligned}
\end{equation}
Here, $\vec{e}_{kk}$ ($k=\mu,h$) is the unit vector along the momentum of the dimuon (dihadron) system in the rest frame of the \Dz meson and $\vec{e}_{k^{+}}$ is the unit vector along the momentum of the positively charged muon (hadron) in the dimuon (dihadron) rest frame. The angle $\phi$ is the angle between the two decay planes of the dimuon and dihadron systems, defined by:    
\begin{equation}
\begin{aligned}   
\cos{\phi} &= \vec{n}_{\mu\mu} \cdot \vec{n}_{hh}, \\
\sin{\phi} &= [\vec{n}_{\mu\mu} \times \vec{n}_{hh}] \cdot \vec{e}_{hh}, 
\end{aligned}
\end{equation}
where $\vec{n}_{kk}=\vec{e}_{k^+}\times\vec{e}_{k^-}$ is defined as the unit vector perpendicular to the decay plane spanned by the two muons (or the two hadrons). The angles are defined in $-1 \leq \cos \theta_h \leq 1$, $ -1 \leq \cos \theta_\mu \leq 1$ and $-\pi \leq \phi  \leq  \pi $. In contrast to Ref.~\cite{LHCb-PAPER-2018-020}, the same definition of the angles is kept for \Dz and \Dzb mesons and the \CP-oddity of angular coefficients is considered in the definition of the \CP averages and \CP asymmetries of the observables. 

The eight coefficients $I_{2-9}$ can be obtained from Eq.~\eqref{supp:eq:diffBF_appendix} by defining piece-wise angular-integration ranges in $\phi$ and $\cos\theta_{\mu}$ as follows:
\begin{equation}
\begin{aligned}   
\label{supp:eq:I2ToI9}
I_2  & = \int_{-\pi}^\pi  d\phi \, \left[ \int_{-1}^{-0.5} d\cos\theta_{\mu} + \int_{0.5}^{1} d\cos\theta_{\mu} - \int_{-0.5}^{0.5}  d\cos\theta_{\mu}   \right]  \, \frac{d^5\Gamma }{dq^2\,dp^2\,d\Vec{\Omega}}\,, \\
I_3  & =\ \frac{3 \pi}{8} \left[ \int_{-\pi}^{-\frac{3\pi}{4}} d \phi  +  \int_{-\frac{\pi}{4}}^{\frac{\pi}{4}} d \phi + \int_{\frac{3\pi}{4}}^\pi d \phi -  \int_{-\frac{3\pi}{4}}^{-\frac{\pi}{4}} d \phi   -  \int_{\frac{\pi}{4}}^{\frac{3\pi}{4}} d \phi    \right] \int_{-1}^1  d\cos\theta_{\mu} \, \frac{d^5\Gamma }{dq^2\,dp^2\,d\Vec{\Omega}}\,,\\
I_4  & =   \frac{3 \pi}{8}   \left[   \int_{-\frac{\pi}{2}}^{\frac{\pi}{2}} d \phi - \int_{-\pi}^{-\frac{\pi}{2}} d \phi   -  \int_{\frac{\pi}{2}}^{\pi} d \phi \right]  \left[  \int_0^1  d\cos\theta_{\mu}  - \int_{-1}^0  d\cos\theta_{\mu}  \right] \, \frac{d^5\Gamma }{dq^2\,dp^2\,d\Vec{\Omega}}\,, \\
I_5  & = \ \left[   \int_{-\frac{\pi}{2}}^{\frac{\pi}{2}} d \phi - \int_{-\pi}^{-\frac{\pi}{2}} d \phi   -  \int_{\frac{\pi}{2}}^{\pi} d \phi \right]  \, \int_{-1}^1  d\cos\theta_{\mu} \, \frac{d^5\Gamma }{dq^2\,dp^2\,d\Vec{\Omega}}\,, \\
I_6 &= \int_{-\pi}^\pi  d\phi \, \left[ \int_0^1 d\cos\theta_{\mu}  - \int_{-1}^0  d\cos\theta_{\mu} \right]\, \frac{d^5\Gamma }{dq^2\,dp^2\,d\Vec{\Omega}}\,, \\
I_7  & =\  \left[ \int_0^{\pi}  d  \phi  - \int_{-\pi}^{0}  d \phi  \right]  \, \int_{-1}^1  d\cos\theta_{\mu} \, \frac{d^5\Gamma }{dq^2\,dp^2\,d\Vec{\Omega}}\,, \\
I_8  & = \ \frac{3 \pi}{8}   \left[ \int_0^{\pi}  d  \phi  - \int_{-\pi}^{0}  d \phi  \right]  \left[  \int_0^1  d\cos\theta_{\mu}  - \int_{-1}^0  d\cos\theta_{\mu}  \right] \, \frac{d^5\Gamma }{dq^2\,dp^2\,d\Vec{\Omega}}\,, \\
I_9  & =  \frac{3 \pi}{8}    \left[\int_{-\pi}^{-\frac{\pi}{2}} d \phi +  \int_{0}^{\frac{\pi}{2}} d \phi - \int_{-\frac{\pi}{2}}^{0} d \phi     -  \int_{\frac{\pi}{2}}^{\pi} d \phi  \right]  \, \int_{-1}^1  d\cos\theta_{\mu} \, \frac{d^5\Gamma }{dq^2\,dp^2\,d\Vec{\Omega}}\,. \\
\end{aligned}
\end{equation}
The normalised and integrated observables $\langle I_i \rangle$ are defined as
\begin{equation}
\begin{aligned}    \label{supp:eq:averagedCoefficients}
 \langle I_{2,3,6,9} \rangle  & = \frac{1}{\Gamma} \int_{q^2_{\min}}^{q_{\max}^2} dq^2 \int_{p_{\min}^2}^{p_{\max}^2} d p^2 \int_{-1}^{+1} d \cos \theta_h \, I_{2,3,6,9} \, , \\
 \langle I_{4,5,7,8} \rangle  & = \frac{1}{\Gamma} \int_{q^2_{\min}}^{q_{\max}^2} dq^2  \int_{p_{\min}^2}^{p_{\max}^2} d p^2  \left[ \int_{0}^{+1} d \cos \theta_h  -  \int_{-1}^{0} d \cos \theta_h  \right] \,I_{4,5,7,8} \, .
\end{aligned}
\end{equation}
Different integration ranges in $q^2$ are defined, according to the expected presence of intermediate hadronic resonances decaying into two muons. The integration in $\cos \theta_h$ is defined to optimise the sensitivity to beyond-SM effects by integrating out contributions from the dominant P-wave resonances in the dihadron system, which further decay into $h^+h^-$({$\rho \to \pip\pim$} and {$\phi \to \Kp\Km$} for \Dppmm and \Dkkmm decays, respectively). No further attempt to separate partial waves contributing to the coefficients $\Ii$ is made.

Experimentally, the observables $\Ii$ are determined by measuring the decay-rate asymmetries of the data split by angular \textit{tags}, $T_i$, defined according to the piece-wise integration of Eq.~\eqref{supp:eq:I2ToI9} and~\eqref{supp:eq:averagedCoefficients}
\begin{equation}
\Ii = \frac{1}{\Gamma} \,a_i\, \bigl[\Gamma(T_i = \texttt{true}) - \Gamma(T_i  = \texttt{false})\bigr]\,, 
\end{equation}
where $a_{2,5,6,7}=1$, $a_{3,4,8,9}={3\pi}/{8}$ and
\begin{equation}
\begin{aligned}
T_2 &= |\cos\theta_\mu|>0.5\,,\\
T_3 &= \cos 2\phi>0\,,\\
T_4 &= (\sin2\theta_{\mu}>0;\,\cos\phi>0;\,\cos\theta_{h}>0)\;\texttt{or}\;(\sin2\theta_{\mu}>0;\, \cos\phi<0;\,\cos\theta_{h}<0)\;\\
    &\texttt{or}\;(\sin2\theta_{\mu}<0;\,\cos \phi<0;\,\cos \theta_{h}>0)\;\texttt{or}\;(\sin2\theta_{\mu}<0;\,\cos\phi>0;\,\cos\theta_{h}<0) \,,\\
T_5 &= (\cos\phi>0;\,\cos\theta_{h}>0) \,\texttt{or}\, (\cos\phi<0;\,\cos\theta_{h}<0)\,,\\
T_6 &= \cos\theta_{\mu}>0\,,\\
T_7 &= (\sin\phi>0;\,\cos\theta_{h}>0)\;\texttt{or}\;(\sin\phi<0;\,\cos\theta_{h}<0)\,,\\
T_8 &= (\sin2\theta_{\mu}>0;\,\sin\phi>0;\,\cos\theta_{h}>0)\;\texttt{or}\;(\sin2\theta_{\mu}>0;\,\sin\phi<0;\,\cos\theta_{h}<0)\;\\
    &\texttt{or}\;(\sin2\theta_{\mu}<0;\,\sin\phi<0;\, \cos\theta_{h}>0)\;\texttt{or}\;(\sin2\theta_{\mu}<0;\,\sin\phi>0;\,\cos\theta_{h}<0)\,,\\
T_9 &= \sin2\phi>0\,.\\
\end{aligned}
\end{equation}

The observables $\Ii$, measured separately for \Dz and \Dzb mesons, are labelled as $\Ii$ and $\Ibi$, respectively. The observables reported in the Letter are the \CP averages, $\Si$, and asymmetries, $\Ai$, defined as
\begin{equation}
\begin{aligned}   
\Si &= \frac{1}{2} \left[ \Ii +(-) \Ibi \right]\,,\\
\Ai &= \frac{1}{2} \left[ \Ii  -(+) \Ibi \right],\, 
\end{aligned}
\end{equation}
for the \CP-even (\CP-odd) coefficients $\langle I_{2,3,4,7} \rangle$ ($\langle I_{5,6,8,9} \rangle$). 

The full set of measured \CP-averaged angular observables for both decay modes is reported in Table~\ref{table:asyS}; their \CP asymmetries are reported in Table~\ref{table:asyA}; the \CP asymmetry in the decay rates \Acp, are reported in Table~\ref{table:ACP}.

\begin{table*}[b]
\centering
\caption{Observable \Acp for (top) \Dppmm and (bottom) \Dkkmm decays in the dimuon-mass regions. The first uncertainty is statistical, the second systematic.}\label{table:ACP}
\begin{tabular}{r@{--}l r@{\,$\pm$\,}c@{\,$\pm$\,}l}
\toprule
\multicolumn{2}{l}{$\mmumu$} & \multicolumn{3}{c}{$\Acp$ [\%]} \\
\multicolumn{2}{l}{[$\mevcc$]} & \multicolumn{3}{c}{} \\
\midrule
\multicolumn{5}{c}{$\Dppmm$} \\
\multicolumn{2}{c}{$<525$} & 28 & 13 & 1 \\ 
525 & 565 & \multicolumn{3}{c}{$-$}\\
565 & 780 & $-$2.7 & 4.1 & 0.4 \\
780 & 950 & $-$1.9 & 5.8 & 0.4 \\ 
950 & 1020 & 0.5 & 3.7 & 0.4 \\ 
1020 & 1100 & 4.2 & 3.4 & 0.4  \\ 
\multicolumn{2}{c}{$>1100$} & \multicolumn{2}{c}{$-$} \\
\multicolumn{2}{c}{Full range} & 2.9 & 2.1 & 0.4 \\ 
\midrule
\multicolumn{5}{c}{$\Dkkmm$} \\
\multicolumn{2}{c}{$<525$} & 4 & 15 & 1\\ 
525 & 565 & \multicolumn{3}{c}{$-$}\\
\multicolumn{2}{c}{$>565$} & $-$2.5 & 6.8 & 0.6  \\ 
\multicolumn{2}{c}{Full range} & $-$2.3 & 6.3 & 0.6 \\ 
\bottomrule
\end{tabular}
\end{table*}

\begin{sidewaystable}[p!]
\centering
\caption{Angular observables $\Si$ for (top) \Dppmm and (bottom) \Dkkmm decays in the different dimuon-mass regions reported in the first column. The first uncertainty is statistical and the second systematic.}\label{table:asyS}
\resizebox{
\ifdim\width>\textwidth
	\textwidth
\else
	\width
\fi}{!}{
\ifthenelse{\boolean{wordcount}}{}{%
\begin{tabular}{r@{--}l r@{\,$\pm$\,}c@{\,$\pm$\,}l r@{\,$\pm$\,}c@{\,$\pm$\,}l r@{\,$\pm$\,}c@{\,$\pm$\,}l r@{\,$\pm$\,}c@{\,$\pm$\,}l r@{\,$\pm$\,}c@{\,$\pm$\,}l r@{\,$\pm$\,}c@{\,$\pm$\,}l r@{\,$\pm$\,}c@{\,$\pm$\,}l r@{\,$\pm$\,}c@{\,$\pm$\,}l }
\toprule
\multicolumn{2}{l}{$\mmumu$} & \multicolumn{3}{c}{$\Sii$ [\%]} & \multicolumn{3}{c}{$\Siii$ [\%]} & \multicolumn{3}{c}{$\Siv$ [\%]} & \multicolumn{3}{c}{$\Sv$ [\%]} & \multicolumn{3}{c}{$\Svi$ [\%]} & \multicolumn{3}{c}{$\Svii$ [\%]} & \multicolumn{3}{c}{$\Sviii$ [\%]} & \multicolumn{3}{c}{$\Six$ [\%]} \\
\multicolumn{2}{l}{[$\mevcc$]} & \multicolumn{3}{c}{} & \multicolumn{3}{c}{} & \multicolumn{3}{c}{} & \multicolumn{3}{c}{} & \multicolumn{3}{c}{} & \multicolumn{3}{c}{} & \multicolumn{3}{c}{} & \multicolumn{3}{c}{}\\
\midrule
\multicolumn{26}{c}{$\Dppmm$} \\
\multicolumn{2}{c}{$<525$} & 5 & 14 & 4 & $-$6 & 16 & 2 & 21 & 16 & 2 & $-$20 & 14 & 1 & $-$14 & 14 & 1 & 8 & 14 & 1 & 16 & 17 & 1 & 26 & 16 & 2 \\ 
525 & 565 & \multicolumn{3}{c}{$-$} & \multicolumn{3}{c}{$-$} & \multicolumn{3}{c}{$-$} & \multicolumn{3}{c}{$-$} & \multicolumn{3}{c}{$-$} & \multicolumn{3}{c}{$-$} & \multicolumn{3}{c}{$-$} & \multicolumn{3}{c}{$-$}\\
565 & 780 & $-$2.4 & 4.1 & 1.1 & $-$9.1 & 4.8 & 1.5 & 3.7 & 4.9 & 1.3 & $-$3.0 & 4.1 & 0.8 & 2.5 & 4.1 & 0.6 & 0.8 & 4.1 & 1.0 & 12.9 & 4.9 & 1.0 & $-$0.1 & 4.9 & 0.9\\
780 & 950 & $-$10.7 & 5.8 & 1.1 & 7.7 & 6.9 & 1.0 & $-$4.7 & 6.9 & 1.5 & 4.7 & 5.8 & 0.7 & 9.0 & 5.8 & 0.7 & $-$4.7 & 5.8 & 1.0 & 1.4 & 6.9 & 0.7 & $-$4.7 & 6.8 & 0.8\\ 
950 & 1020 & $-$2.0 & 3.7 & 1.6 & $-$17.4 & 4.3 & 1.5 & $-$9.9 & 4.3 & 3.5 & 2.0 & 3.7 & 0.8 & 6.5 & 3.7 & 1.4 & $-$3.6 & 3.7 & 1.1 & 2.6 & 4.3 & 0.9 & 16.9 & 4.3 & 1.0\\ 
1020 & 1100 & 1.7 & 3.4 & 1.5 & $-$15.3 & 4.0 & 1.7 & $-$18.3 & 4.0 & 2.5 & $-$6.9 & 3.4 & 1.2 & 1.1 & 3.4 & 0.8 & 2.7 & 3.4 & 2.0 & 0.7 & 4.1 & 0.9 & 7.8 & 4.0 & 1.7\\ 
\multicolumn{2}{c}{$>1100$} & \multicolumn{2}{c}{$-$} & \multicolumn{2}{c}{$-$} & \multicolumn{2}{c}{$-$} & \multicolumn{3}{c}{$-$} & \multicolumn{3}{c}{$-$} & \multicolumn{3}{c}{$-$} & \multicolumn{3}{c}{$-$} & \multicolumn{3}{c}{$-$}\\
\multicolumn{2}{c}{Full range} & $-$3.4 & 2.1 & 1.0 & $-$10.4 & 2.5 & 0.9 & $-$4.6 & 2.5 & 1.6 & $-$2.9 & 2.1 & 0.6 & 3.7 & 2.1 & 0.5 & $-$0.6 & 2.1 & 0.9 & 3.8 & 2.5 & 0.5 & 5.1 & 2.5 & 0.5\\ 
\midrule
\multicolumn{26}{c}{$\Dkkmm$} \\
\multicolumn{2}{c}{$<525$} & $-$2 & 15 & 2 & $-$20 & 17 & 3 & 17 & 18 & 2 & $-$35 & 14 & 1 & $-$6 & 15 & 2 & $-$4 & 15 & 1 & 9 & 18 & 1 & 22 & 16 & 1\\ 
525 & 565 & \multicolumn{3}{c}{$-$} & \multicolumn{3}{c}{$-$} & \multicolumn{3}{c}{$-$} &  \multicolumn{3}{c}{$-$} & \multicolumn{3}{c}{$-$} & \multicolumn{3}{c}{$-$}& \multicolumn{3}{c}{$-$}& \multicolumn{3}{c}{$-$}\\
\multicolumn{2}{c}{$>565$} & 15.9 & 6.6 & 1.5 & $-$18.9 & 8.0 & 1.3 & $-$18.3 & 7.9 & 1.5 & $-$7.5 & 6.8 & 0.6 & 7.8 & 6.8 & 0.5 & 1.7 & 6.8 & 1.5 & 4.0 & 8.0 & 0.5 & 9.3 & 8.0 & 0.6\\ 
\multicolumn{2}{c}{Full range} & 12.4 & 6.1 & 1.7 & $-$17.5 & 7.4 & 1.3 & $-$12.5 & 7.3 & 1.8 & $-$12.9 & 6.2 & 0.8 & 5.7 & 6.2 & 0.8 & 1.7 & 6.3 & 1.1 & 3.4 & 7.4 & 1.0 & 12.0 & 7.3 & 0.8\\ 
\bottomrule
\end{tabular}
}}
\end{sidewaystable}

\begin{sidewaystable}[p!]
\centering
\caption{Angular observables $\Ai$ for (top) \Dppmm and (bottom) \Dkkmm decays in the different dimuon-mass regions reported in the first column. The first uncertainty is statistical and the second systematic.}\label{table:asyA}
\resizebox{
\ifdim\width>\textwidth
	\textwidth
\else
	\width
\fi}{!}{
\ifthenelse{\boolean{wordcount}}{}{%
\begin{tabular}{r@{--}l r@{\,$\pm$\,}c@{\,$\pm$\,}l r@{\,$\pm$\,}c@{\,$\pm$\,}l r@{\,$\pm$\,}c@{\,$\pm$\,}l r@{\,$\pm$\,}c@{\,$\pm$\,}l r@{\,$\pm$\,}c@{\,$\pm$\,}l r@{\,$\pm$\,}c@{\,$\pm$\,}l r@{\,$\pm$\,}c@{\,$\pm$\,}l r@{\,$\pm$\,}c@{\,$\pm$\,}l }
\toprule
\multicolumn{2}{l}{$\mmumu$} & \multicolumn{3}{c}{$\Aii$ [\%]} & \multicolumn{3}{c}{$\Aiii$ [\%]} & \multicolumn{3}{c}{$\Aiv$ [\%]} & \multicolumn{3}{c}{$\Av$ [\%]} & \multicolumn{3}{c}{$\Avi$ [\%]} & \multicolumn{3}{c}{$\Avii$ [\%]} & \multicolumn{3}{c}{$\Aviii$ [\%]} & \multicolumn{3}{c}{$\Aix$ [\%]} \\
\multicolumn{2}{l}{[$\mevcc$]} & \multicolumn{3}{c}{} & \multicolumn{3}{c}{} & \multicolumn{3}{c}{} & \multicolumn{3}{c}{} & \multicolumn{3}{c}{} & \multicolumn{3}{c}{} & \multicolumn{3}{c}{} & \multicolumn{3}{c}{}\\
\midrule
\multicolumn{26}{c}{$\Dppmm$} \\
\multicolumn{2}{c}{$<525$} & $-$10 & 14 & 2 & 2 & 16 & 1 & $-$7 & 16 & 2 & 16 & 14 & 1 & 0 & 14 & 1 & $-$10 & 14 & 2 & 3 & 17 & 2 & $-$25 & 16 & 2 \\ 
525 & 565 & \multicolumn{3}{c}{$-$} & \multicolumn{3}{c}{$-$} & \multicolumn{3}{c}{$-$} & \multicolumn{3}{c}{$-$} & \multicolumn{3}{c}{$-$} & \multicolumn{3}{c}{$-$} & \multicolumn{3}{c}{$-$} & \multicolumn{3}{c}{$-$}\\
565 & 780 & $-$1.1 & 4.1 & 1.9 & 5.7 & 4.8 & 0.7 & 0.6 & 4.9 & 0.7 & $-$3.0 & 4.1 & 1.1 & $-$4.8 & 4.1 & 1.0 & $-$3.5 & 4.1 & 1.0 & $-$1.8 & 4.9 & 1.2 & 1.6 & 4.9 & 1.1\\
780 & 950 & $-$7.7 & 5.8 & 0.6 & 3.9 & 6.9 & 0.8 & 1.2 & 6.9 & 0.7 & $-$3.3 & 5.8 & 1.0 & 0.4 & 5.8 & 1.0 & $-$2.6 & 5.8 & 0.6 & $-$5.1 & 6.9 & 1.5 & $-$2.9 & 6.8 & 1.0\\ 
950 & 1020 & 2.3 & 3.7 & 0.7 & $-$2.2 & 4.3 & 2.1 & 7.6 & 4.3 & 0.9 & $-$3.6 & 3.7 & 1.2 & 4.5 & 3.7 & 1.1 & 3.5 & 3.7 & 0.9 & 2.7 & 4.3 & 1.3 & 1.4 & 4.3 & 1.2\\ 
1020 & 1100 & $-$4.8 & 3.4 & 0.9 & $-$2.6 & 4.0 & 1.2 & $-$2.4 & 4.0 & 1.0 & $-$2.3 & 3.4 & 1.2 & 3.2 & 3.4 & 1.1 & $-$1.3 & 3.4 & 0.8 & 5.1 & 4.1 & $-$1.3 & $-$5.9 & 4.0 & 1.8\\ 
\multicolumn{2}{c}{$>1100$} & \multicolumn{2}{c}{$-$} & \multicolumn{2}{c}{$-$} & \multicolumn{2}{c}{$-$} & \multicolumn{3}{c}{$-$} & \multicolumn{3}{c}{$-$} & \multicolumn{3}{c}{$-$} & \multicolumn{3}{c}{$-$} & \multicolumn{3}{c}{$-$}\\
\multicolumn{2}{c}{Full range} & $-$2.6 & 2.1 & 0.7 & 0.9 & 2.5 & 0.5 & 1.3 & 2.5 & 0.5 & $-$1.7 & 2.1 & 0.9 & $-$0.5 & 2.1 & 0.9 & $-$1.5 & 2.1 & 0.6 & 0.3 & 2.5 & 1.0 & $-$2.1 & 2.5 & 0.9\\ 
\midrule
\multicolumn{26}{c}{$\Dkkmm$} \\
\multicolumn{2}{c}{$<525$} & $-$31 & 15 & 2 & 41 & 17 & 2 & $-$23 & 18 & 3 & 13 & 14 & 1 & 20 & 15 & 2 & $-$4 & 15 & 1 & $-$15 & 18 & 2 & 45 & 16 & 2\\ 
525 & 565 & \multicolumn{3}{c}{$-$} & \multicolumn{3}{c}{$-$} & \multicolumn{3}{c}{$-$} &  \multicolumn{3}{c}{$-$} & \multicolumn{3}{c}{$-$} & \multicolumn{3}{c}{$-$}& \multicolumn{3}{c}{$-$}& \multicolumn{3}{c}{$-$}\\
\multicolumn{2}{c}{$>565$} & 2.6 & 6.6 & 0.5 & 2.8 & 8.0 & 0.5 & $-$10.1 & 7.9 & 0.6 & 0.2 & 6.8 & 1.3 & $-$1.2 & 6.8 & 0.5 & 4.7 & 6.8 & 0.5 & $-$6.9 & 8.0 & 1.4 & 1.2 & 8.0 & 1.2\\ 
\multicolumn{2}{c}{Full range} & $-$4.1 & 6.1 & 0.9 & 9.3 & 7.4 & 0.8 & $-$12.4 & 7.3 & 1.3 & 2.5 & 6.2 & 1.1 & 0.1 & 6.2 & 1.1 & 2.9 & 6.3 & 0.9 & $-$6.8 & 7.4 & 1.2 & 11.2 & 7.3 & 1.2\\ 
\bottomrule
\end{tabular}
}}
\end{sidewaystable}

Observables measured in disjoint dimuon-mass intervals are statistically uncorrelated. The statistical correlations between the observables for a given dimuon-mass interval for \Dppmm and \Dkkmm decays are reported in Tables~\ref{table:corr_pipi_S_A_bin0}--\ref{table:corr_pipi_S_A_bin6} and \ref{table:corr_KK_S_A_bin0}--\ref{table:corr_KK_S_A_bin3}, respectively. Systematic uncertainties between the observables, in each and across different dimuon-mass intervals, are assumed to be fully correlated. 

\begin{table}[ht]
\centering 
\caption{Correlation matrix for the observables $\Acp$, $\langle S_i \rangle$ and $\langle A_i \rangle$ for \Dppmm decays measured in the dimuon-mass-integrated interval.}
\label{table:corr_pipi_S_A_bin0}
\resizebox{1.0\textwidth}{!}{  
\begin{tabular}[c]{crrrrrrrrrrrrrrrrrr}
\toprule
  	&    $\Acp$  &  $\langle S_2 \rangle$&$\langle S_3 \rangle$&$\langle S_4 \rangle$&$\langle S_5 \rangle$&$\langle S_6 \rangle$&$\langle S_7 \rangle$&$\langle S_8 \rangle$  & $\langle S_9 \rangle$ &  $\langle A_2 \rangle$&$\langle A_3 \rangle$&$\langle A_4 \rangle$&$\langle A_5 \rangle$&$\langle A_6 \rangle$&$\langle A_7 \rangle$&$\langle A_8 \rangle$  & $\langle A_9 \rangle$ \\
$\Acp$& 1.00 &  $-$0.04 &  $-$0.04 &  0.01 &  0.04  & 0.01 &  $-$0.09  & $-$0.08  & $-$0.03 &  0.05 &  $-$0.07 &  0.02 &  0.00  & $-$0.06 &  0.06  & 0.01  & $-$0.08  \\ 
$\langle S_2 \rangle$&           &  1.00 &  0.02 &  $-$0.03 &  $-$0.06  & 0.08 &  $-$0.04  & $-$0.02  & 0.01  & $-$0.09 &  $-$0.03 &  $-$0.01 &  0.07  & $-$0.08 &  $-$0.04  & $-$0.11  & 0.02  \\ 
$\langle S_3 \rangle$&           &             &  	 1.00 &  $-$0.04 &  $-$0.06  & $-$0.06 &  $-$0.06  & 0.05  & 0.05  & 0.00 &  0.01 &  0.09 &  0.01  & $-$0.01 &  0.01  & $-$0.02  & 0.11  \\ 
$\langle S_4 \rangle$&           &             &               &  1.00 &  0.02  & $-$0.05 &  $-$0.00  & 0.07  & 0.03  & $-$0.03 &  0.06 &  $-$0.02 &  $-$0.05  & 0.00 &  $-$0.00  & $-$0.06  & $-$0.09  \\ 
$\langle S_5 \rangle$&           &             &               &             &  1.00  & 0.01 &  0.05  & $-$0.05  & 0.07  & 0.02 &  0.04 &  $-$0.05 &  $-$0.07  & 0.01 &  $-$0.04  & $-$0.01  & $-$0.05  \\ 
$\langle S_6 \rangle$&           &             &               &             &              & 1.00 &  0.04  & 0.03  & $-$0.01  & $-$0.06 &  $-$0.06 &  0.03 &  0.04  & $-$0.03 &  $-$0.03  & $-$0.00  & $-$0.00  \\ 
$\langle S_7 \rangle$&           &             &               &             &              &            &  1.00  & $-$0.02  & $-$0.08  & 0.01 &  0.05 &  $-$0.07 &  0.01  & 0.01 &  $-$0.01  & $-$0.07  & 0.00  \\ 
$\langle S_8 \rangle$&           &             &               &             &              &            &              & 1.00  & 0.01  & 0.01 &  $-$0.02 &  0.02 &  0.03  & 0.04 &  $-$0.05  & $-$0.03  & $-$0.01  \\ 
$\langle S_9 \rangle$&           &             &               &             &              &            &              & 		    & 1.00  & $-$0.05 &  0.10 &  0.03 &  $-$0.01  & $-$0.03 &  0.00  & $-$0.00  & $-$0.01  \\ 
$\langle A_2 \rangle$&           &             &               &             &              &            &              & 		    & 		    & 1.00 &  $-$0.10 &  0.01 &  $-$0.02  & 0.08 &  0.06  & 0.10  & 0.06  \\ 
$\langle A_3 \rangle$&           &             &               &             &              &            &              & 		    & 		    & 		   &  1.00 &  0.00 &  0.05  & $-$0.08 &  0.00  & $-$0.01  & $-$0.00  \\ 
$\langle A_4 \rangle$&           &             &               &             &              &            &              & 		    & 		    & 		   &  		   &  1.00 &  0.03  & $-$0.01 &  $-$0.03  & 0.02  & 0.02  \\ 
$\langle A_5 \rangle$&           &             &               &             &              &            &              & 		    & 		    & 		   &  		   &  		   &  1.00  & $-$0.02 &  0.13  & $-$0.04  & 0.02  \\ 
$\langle A_6 \rangle$&           &             &               &             &              &            &              & 		    & 		    & 		   &  		   &  		   &  		    & 1.00 &  0.03  & 0.02  & $-$0.07  \\ 
$\langle A_7 \rangle$&           &             &               &             &              &            &              & 		    & 		    & 		   &  		   &  		   &  		    & 		   &  1.00  & 0.06  & $-$0.02  \\ 
$\langle A_8 \rangle$&           &             &               &             &              &            &              & 		    & 		    & 		   &  		   &  		   &  		    & 		   &  		    & 1.00  & $-$0.09  \\ 
$\langle A_9 \rangle$&           &             &               &             &              &            &              & 		    & 		    & 		   &  		   &  		   &  		    & 		   &  		    & 		    & 1.00  \\ 
\bottomrule
\end{tabular}}
\end{table}

\begin{table}[ht]
\centering 
\caption{Correlation matrix for the observables $\Acp$, $\langle S_i \rangle$ and $\langle A_i \rangle$ for \Dppmm decays measured in the interval \mbox{$m(\mu^+\mu^-) <525 \mevcc$}.}
\label{table:corr_pipi_S_A_bin1}
 \resizebox{1.0\textwidth}{!}{  
\begin{tabular}[c]{crrrrrrrrrrrrrrrrrr}
\toprule
  	&    $\Acp$  &  $\langle S_2 \rangle$&$\langle S_3 \rangle$&$\langle S_4 \rangle$&$\langle S_5 \rangle$&$\langle S_6 \rangle$&$\langle S_7 \rangle$&$\langle S_8 \rangle$  & $\langle S_9 \rangle$ &  $\langle A_2 \rangle$&$\langle A_3 \rangle$&$\langle A_4 \rangle$&$\langle A_5 \rangle$&$\langle A_6 \rangle$&$\langle A_7 \rangle$&$\langle A_8 \rangle$  & $\langle A_9 \rangle$ \\
$\Acp$& 1.00 &  0.01 &  0.06 &  $-$0.01 &  $-$0.09  & $-$0.09 &  0.12  & $-$0.16  & 0.04 &  0.11 &  $-$0.07 &  $-$0.08 &  0.19  & 0.12 &  $-$0.07  & 0.16  & $-$0.07  \\ 
$\langle S_2 \rangle$&           &  1.00 &  $-$0.02 &  0.03 &  0.10  & 0.19 &  $-$0.13  & 0.12  & 0.23  & $-$0.45 &  0.03 &  $-$0.05 &  0.11  & $-$0.18 &  0.07  & $-$0.24  & $-$0.21  \\ 
$\langle S_3 \rangle$&           &             &  	 1.00 &  $-$0.01 &  $-$0.07  & $-$0.21 &  0.07  & $-$0.10  & $-$0.16  & 0.08 &  $-$0.37 &  0.15 &  $-$0.07  & 0.22 &  $-$0.11  & 0.05  & 0.07  \\ 
$\langle S_4 \rangle$&           &             &               &  1.00 &  0.01  & $-$0.06 &  0.09  & 0.19  & $-$0.10  & 0.00 &  0.10 &  $-$0.40 &  $-$0.11  & 0.08 &  $-$0.10  & $-$0.25  & 0.12  \\ 
$\langle S_5 \rangle$&           &             &               &             &  1.00  & 0.21 &  0.17  & 0.11  & 0.01  & 0.09 &  $-$0.08 &  $-$0.10 &  $-$0.39  & $-$0.21 &  $-$0.21  & $-$0.13  & $-$0.10  \\ 
$\langle S_6 \rangle$&           &             &               &             &              & 1.00 &  $-$0.17  & 0.02  & 0.13  & $-$0.20 &  0.18 &  0.12 &  $-$0.19  & $-$0.43 &  0.04  & $-$0.09  & $-$0.18  \\ 
$\langle S_7 \rangle$&           &             &               &             &              &            &  1.00  & $-$0.04  & $-$0.17  & 0.12 &  $-$0.09 &  $-$0.12 &  $-$0.13  & 0.10 &  $-$0.44  & 0.04  & 0.20  \\ 
$\langle S_8 \rangle$&           &             &               &             &              &            &              & 1.00  & 0.22  & $-$0.18 &  0.02 &  $-$0.20 &  $-$0.11  & $-$0.16 &  $-$0.02  & $-$0.39  & $-$0.14  \\ 
$\langle S_9 \rangle$&           &             &               &             &              &            &              & 		    & 1.00  & $-$0.23 &  0.09 &  $-$0.01 &  $-$0.14  & $-$0.15 &  0.15  & $-$0.18  & $-$0.38  \\ 
$\langle A_2 \rangle$&           &             &               &             &              &            &              & 		    & 		    & 1.00 &  $-$0.04 &  $-$0.05 &  0.14  & 0.27 &  $-$0.12  & 0.13  & 0.27  \\ 
$\langle A_3 \rangle$&           &             &               &             &              &            &              & 		    & 		    & 		   &  1.00 &  $-$0.04 &  $-$0.10  & $-$0.13 &  0.10  & $-$0.09  & $-$0.09  \\ 
$\langle A_4 \rangle$&           &             &               &             &              &            &              & 		    & 		    & 		   &  		   &  1.00 &  $-$0.05  & $-$0.06 &  0.19  & 0.18  & $-$0.14  \\ 
$\langle A_5 \rangle$&           &             &               &             &              &            &              & 		    & 		    & 		   &  		   &  		   &  1.00  & 0.24 &  0.20  & 0.11  & 0.07  \\ 
$\langle A_6 \rangle$&           &             &               &             &              &            &              & 		    & 		    & 		   &  		   &  		   &  		    & 1.00 &  $-$0.08  & 0.04  & 0.21  \\ 
$\langle A_7 \rangle$&           &             &               &             &              &            &              & 		    & 		    & 		   &  		   &  		   &  		    & 		   &  1.00  & $-$0.09  & $-$0.10  \\ 
$\langle A_8 \rangle$&           &             &               &             &              &            &              & 		    & 		    & 		   &  		   &  		   &  		    & 		   &  		    & 1.00  & 0.21  \\ 
$\langle A_9 \rangle$&           &             &               &             &              &            &              & 		    & 		    & 		   &  		   &  		   &  		    & 		   &  		    & 		    & 1.00  \\ 
\bottomrule
\end{tabular}}
\end{table}

\begin{table}[ht]
\centering 
\caption{Correlation matrix for the observables $\Acp$, $\langle S_i \rangle$ and $\langle A_i \rangle$ for \Dppmm decays measured in the interval \mbox{$ 565 < m(\mu^+\mu^-) <780 \mevcc$}.}
\label{table:corr_pipi_S_A_bin3}
 \resizebox{1.0\textwidth}{!}{  
\begin{tabular}[c]{crrrrrrrrrrrrrrrrrr}
\toprule
  	&    $\Acp$  &  $\langle S_2 \rangle$&$\langle S_3 \rangle$&$\langle S_4 \rangle$&$\langle S_5 \rangle$&$\langle S_6 \rangle$&$\langle S_7 \rangle$&$\langle S_8 \rangle$  & $\langle S_9 \rangle$ &  $\langle A_2 \rangle$&$\langle A_3 \rangle$&$\langle A_4 \rangle$&$\langle A_5 \rangle$&$\langle A_6 \rangle$&$\langle A_7 \rangle$&$\langle A_8 \rangle$  & $\langle A_9 \rangle$ \\
$\Acp$& 1.00 &  0.02 &  $-$0.01 &  0.05 &  0.01  & $-$0.02 &  $-$0.03  & $-$0.04  & $-$0.01 &  0.12 &  $-$0.00 &  0.04 &  0.00  & $-$0.08 &  $-$0.01  & $-$0.05  & 0.03  \\ 
$\langle S_2 \rangle$&           &  1.00 &  $-$0.04 &  0.06 &  $-$0.04  & $-$0.02 &  $-$0.04  & 0.01  & 0.12  & 0.01 &  0.01 &  $-$0.03 &  0.04  & $-$0.10 &  $-$0.16  & $-$0.01  & $-$0.01  \\ 
$\langle S_3 \rangle$&           &             &  	 1.00 &  $-$0.07 &  $-$0.00  & $-$0.05 &  0.05  & $-$0.07  & 0.03  & 0.05 &  0.05 &  0.06 &  $-$0.04  & $-$0.04 &  $-$0.04  & 0.04  & 0.08  \\ 
$\langle S_4 \rangle$&           &             &               &  1.00 &  $-$0.04  & $-$0.05 &  $-$0.07  & 0.03  & 0.14  & $-$0.02 &  0.02 &  0.07 &  $-$0.06  & $-$0.02 &  $-$0.04  & 0.07  & $-$0.07  \\ 
$\langle S_5 \rangle$&           &             &               &             &  1.00  & 0.09 &  $-$0.05  & $-$0.05  & 0.05  & 0.04 &  0.02 &  $-$0.05 &  0.03  & 0.08 &  0.05  & 0.03  & $-$0.03  \\ 
$\langle S_6 \rangle$&           &             &               &             &              & 1.00 &  0.06  & 0.02  & $-$0.04  & $-$0.11 &  $-$0.07 &  $-$0.04 &  $-$0.02  & 0.00 &  $-$0.05  & $-$0.09  & 0.01  \\ 
$\langle S_7 \rangle$&           &             &               &             &              &            &  1.00  & $-$0.01  & $-$0.05  & $-$0.09 &  $-$0.04 &  $-$0.00 &  $-$0.04  & $-$0.08 &  0.03  & $-$0.05  & $-$0.07  \\ 
$\langle S_8 \rangle$&           &             &               &             &              &            &              & 1.00  & 0.04  & $-$0.02 &  0.03 &  0.04 &  $-$0.03  & $-$0.02 &  $-$0.09  & 0.00  & $-$0.03  \\ 
$\langle S_9 \rangle$&           &             &               &             &              &            &              & 		    & 1.00  & $-$0.01 &  0.09 &  $-$0.06 &  $-$0.02  & $-$0.06 &  $-$0.02  & 0.01  & $-$0.02  \\ 
$\langle A_2 \rangle$&           &             &               &             &              &            &              & 		    & 		    & 1.00 &  $-$0.04 &  0.04 &  $-$0.05  & $-$0.04 &  0.05  & 0.01  & 0.06  \\ 
$\langle A_3 \rangle$&           &             &               &             &              &            &              & 		    & 		    & 		   &  1.00 &  0.04 &  0.00  & $-$0.04 &  0.03  & 0.05  & 0.07  \\ 
$\langle A_4 \rangle$&           &             &               &             &              &            &              & 		    & 		    & 		   &  		   &  1.00 &  $-$0.01  & $-$0.10 &  $-$0.03  & 0.11  & 0.11  \\ 
$\langle A_5 \rangle$&           &             &               &             &              &            &              & 		    & 		    & 		   &  		   &  		   &  1.00  & 0.08 &  $-$0.02  & $-$0.10  & 0.04  \\ 
$\langle A_6 \rangle$&           &             &               &             &              &            &              & 		    & 		    & 		   &  		   &  		   &  		    & 1.00 &  0.14  & $-$0.03  & $-$0.05  \\ 
$\langle A_7 \rangle$&           &             &               &             &              &            &              & 		    & 		    & 		   &  		   &  		   &  		    & 		   &  1.00  & $-$0.01  & $-$0.04  \\ 
$\langle A_8 \rangle$&           &             &               &             &              &            &              & 		    & 		    & 		   &  		   &  		   &  		    & 		   &  		    & 1.00  & 0.06  \\ 
$\langle A_9 \rangle$&           &             &               &             &              &            &              & 		    & 		    & 		   &  		   &  		   &  		    & 		   &  		    & 		    & 1.00  \\ 
\bottomrule
\end{tabular}}
\end{table}

\begin{table}[ht]
\centering 
\caption{Correlation matrix for the observables $\Acp$, $\langle S_i \rangle$ and $\langle A_i \rangle$ for \Dppmm decays measured in the interval \mbox{$780 < m(\mu^+\mu^-) <950 \mevcc$}.}
\label{table:corr_pipi_S_A_bin4}
 \resizebox{1.0\textwidth}{!}{  
\begin{tabular}[c]{crrrrrrrrrrrrrrrrrr}
\toprule
  	&    $\Acp$  &  $\langle S_2 \rangle$&$\langle S_3 \rangle$&$\langle S_4 \rangle$&$\langle S_5 \rangle$&$\langle S_6 \rangle$&$\langle S_7 \rangle$&$\langle S_8 \rangle$  & $\langle S_9 \rangle$ &  $\langle A_2 \rangle$&$\langle A_3 \rangle$&$\langle A_4 \rangle$&$\langle A_5 \rangle$&$\langle A_6 \rangle$&$\langle A_7 \rangle$&$\langle A_8 \rangle$  & $\langle A_9 \rangle$ \\
$\Acp$& 1.00 &  $-$0.08 &  0.01 &  0.11 &  0.04  & $-$0.03 &  $-$0.03  & $-$0.05  & $-$0.00 &  0.10 &  $-$0.03 &  0.02 &  $-$0.03  & $-$0.14 &  $-$0.02  & $-$0.07  & 0.09  \\ 
$\langle S_2 \rangle$&           &  1.00 &  $-$0.10 &  $-$0.06 &  $-$0.11  & $-$0.05 &  0.09  & 0.04  & 0.04  & $-$0.01 &  $-$0.09 &  0.08 &  0.04  & $-$0.00 &  $-$0.08  & $-$0.03  & $-$0.03  \\ 
$\langle S_3 \rangle$&           &             &  	 1.00 &  $-$0.02 &  $-$0.09  & 0.01 &  0.09  & 0.11  & 0.04  & $-$0.05 &  0.02 &  $-$0.06 &  0.03  & $-$0.06 &  $-$0.03  & 0.08  & $-$0.06  \\ 
$\langle S_4 \rangle$&           &             &               &  1.00 &  0.01  & 0.04 &  $-$0.13  & 0.03  & 0.02  & 0.17 &  $-$0.02 &  0.02 &  0.01  & $-$0.06 &  $-$0.09  & $-$0.09  & $-$0.02  \\ 
$\langle S_5 \rangle$&           &             &               &             &  1.00  & 0.05 &  $-$0.06  & $-$0.17  & $-$0.07  & 0.07 &  0.04 &  $-$0.01 &  0.02  & 0.07 &  $-$0.11  & $-$0.04  & $-$0.06  \\ 
$\langle S_6 \rangle$&           &             &               &             &              & 1.00 &  0.09  & $-$0.06  & $-$0.03  & $-$0.03 &  0.02 &  $-$0.04 &  0.03  & $-$0.01 &  0.03  & $-$0.08  & 0.02  \\ 
$\langle S_7 \rangle$&           &             &               &             &              &            &  1.00  & $-$0.03  & $-$0.01  & $-$0.02 &  0.01 &  $-$0.11 &  $-$0.04  & $-$0.01 &  $-$0.04  & $-$0.04  & $-$0.06  \\ 
$\langle S_8 \rangle$&           &             &               &             &              &            &              & 1.00  & 0.02  & 0.04 &  0.04 &  $-$0.01 &  $-$0.06  & $-$0.12 &  $-$0.01  & 0.02  & 0.05  \\ 
$\langle S_9 \rangle$&           &             &               &             &              &            &              & 		    & 1.00  & 0.05 &  0.04 &  $-$0.08 &  $-$0.05  & $-$0.04 &  0.05  & 0.11  & $-$0.04  \\ 
$\langle A_2 \rangle$&           &             &               &             &              &            &              & 		    & 		    & 1.00 &  0.02 &  $-$0.05 &  $-$0.00  & $-$0.06 &  0.07  & 0.04  & $-$0.02  \\ 
$\langle A_3 \rangle$&           &             &               &             &              &            &              & 		    & 		    & 		   &  1.00 &  $-$0.14 &  $-$0.03  & $-$0.01 &  0.10  & 0.23  & $-$0.03  \\ 
$\langle A_4 \rangle$&           &             &               &             &              &            &              & 		    & 		    & 		   &  		   &  1.00 &  0.07  & 0.01 &  $-$0.11  & $-$0.02  & 0.03  \\ 
$\langle A_5 \rangle$&           &             &               &             &              &            &              & 		    & 		    & 		   &  		   &  		   &  1.00  & 0.03 &  0.01  & $-$0.08  & $-$0.07  \\ 
$\langle A_6 \rangle$&           &             &               &             &              &            &              & 		    & 		    & 		   &  		   &  		   &  		    & 1.00 &  $-$0.00  & $-$0.07  & $-$0.13  \\ 
$\langle A_7 \rangle$&           &             &               &             &              &            &              & 		    & 		    & 		   &  		   &  		   &  		    & 		   &  1.00  & $-$0.05  & 0.00  \\ 
$\langle A_8 \rangle$&           &             &               &             &              &            &              & 		    & 		    & 		   &  		   &  		   &  		    & 		   &  		    & 1.00  & $-$0.03  \\ 
$\langle A_9 \rangle$&           &             &               &             &              &            &              & 		    & 		    & 		   &  		   &  		   &  		    & 		   &  		    & 		    & 1.00  \\ 
\bottomrule
\end{tabular}}
\end{table}

\begin{table}[ht]
\centering 
\caption{Correlation matrix for the observables $\Acp$, $\langle S_i \rangle$ and $\langle A_i \rangle$ for \Dppmm decays measured in the interval \mbox{$950 < m(\mu^+\mu^-) <1020 \mevcc$}.}
\label{table:corr_pipi_S_A_bin5}
 \resizebox{1.0\textwidth}{!}{  
\begin{tabular}[c]{crrrrrrrrrrrrrrrrrr}
\toprule
  	&    $\Acp$  &  $\langle S_2 \rangle$&$\langle S_3 \rangle$&$\langle S_4 \rangle$&$\langle S_5 \rangle$&$\langle S_6 \rangle$&$\langle S_7 \rangle$&$\langle S_8 \rangle$  & $\langle S_9 \rangle$ &  $\langle A_2 \rangle$&$\langle A_3 \rangle$&$\langle A_4 \rangle$&$\langle A_5 \rangle$&$\langle A_6 \rangle$&$\langle A_7 \rangle$&$\langle A_8 \rangle$  & $\langle A_9 \rangle$ \\
$\Acp$& 1.00 &  $-$0.03 &  0.03 &  $-$0.07 &  0.01  & $-$0.03 &  $-$0.04  & $-$0.03  & $-$0.03 &  0.01 &  $-$0.08 &  $-$0.00 &  $-$0.07  & 0.01 &  0.00  & $-$0.04  & $-$0.06  \\ 
$\langle S_2 \rangle$&           &  1.00 &  $-$0.03 &  $-$0.05 &  $-$0.03  & 0.08 &  0.02  & 0.05  & $-$0.08  & 0.01 &  $-$0.04 &  $-$0.06 &  0.15  & 0.03 &  0.06  & $-$0.01  & $-$0.01  \\ 
$\langle S_3 \rangle$&           &             &  	 1.00 &  $-$0.12 &  $-$0.00  & $-$0.01 &  $-$0.01  & $-$0.04  & 0.14  & $-$0.07 &  $-$0.05 &  0.03 &  $-$0.00  & $-$0.03 &  $-$0.01  & 0.08  & $-$0.04  \\ 
$\langle S_4 \rangle$&           &             &               &  1.00 &  0.02  & $-$0.01 &  0.00  & 0.06  & 0.07  & $-$0.10 &  $-$0.01 &  $-$0.05 &  0.06  & $-$0.05 &  $-$0.01  & $-$0.05  & $-$0.04  \\ 
$\langle S_5 \rangle$&           &             &               &             &  1.00  & $-$0.03 &  0.15  & 0.03  & $-$0.04  & 0.08 &  $-$0.08 &  $-$0.02 &  0.03  & 0.06 &  $-$0.06  & $-$0.06  & 0.02  \\ 
$\langle S_6 \rangle$&           &             &               &             &              & 1.00 &  $-$0.01  & $-$0.01  & 0.02  & 0.01 &  $-$0.07 &  $-$0.01 &  0.06  & 0.02 &  0.02  & $-$0.04  & $-$0.09  \\ 
$\langle S_7 \rangle$&           &             &               &             &              &            &  1.00  & 0.11  & 0.00  & 0.03 &  $-$0.01 &  0.02 &  $-$0.02  & $-$0.04 &  0.00  & $-$0.04  & $-$0.09  \\ 
$\langle S_8 \rangle$&           &             &               &             &              &            &              & 1.00  & $-$0.07  & 0.04 &  0.00 &  $-$0.02 &  $-$0.03  & 0.04 &  $-$0.03  & $-$0.07  & 0.09  \\ 
$\langle S_9 \rangle$&           &             &               &             &              &            &              & 		    & 1.00  & $-$0.04 &  $-$0.02 &  $-$0.07 &  $-$0.03  & $-$0.05 &  $-$0.01  & 0.04  & $-$0.01  \\ 
$\langle A_2 \rangle$&           &             &               &             &              &            &              & 		    & 		    & 1.00 &  $-$0.04 &  $-$0.03 &  $-$0.01  & 0.09 &  0.03  & 0.06  & 0.00  \\ 
$\langle A_3 \rangle$&           &             &               &             &              &            &              & 		    & 		    & 		   &  1.00 &  $-$0.08 &  $-$0.08  & $-$0.02 &  $-$0.07  & $-$0.08  & 0.14  \\ 
$\langle A_4 \rangle$&           &             &               &             &              &            &              & 		    & 		    & 		   &  		   &  1.00 &  0.00  & $-$0.03 &  $-$0.05  & 0.10  & 0.01  \\ 
$\langle A_5 \rangle$&           &             &               &             &              &            &              & 		    & 		    & 		   &  		   &  		   &  1.00  & $-$0.06 &  0.13  & 0.05  & $-$0.01  \\ 
$\langle A_6 \rangle$&           &             &               &             &              &            &              & 		    & 		    & 		   &  		   &  		   &  		    & 1.00 &  0.03  & $-$0.03  & 0.11  \\ 
$\langle A_7 \rangle$&           &             &               &             &              &            &              & 		    & 		    & 		   &  		   &  		   &  		    & 		   &  1.00  & 0.06  & $-$0.03  \\ 
$\langle A_8 \rangle$&           &             &               &             &              &            &              & 		    & 		    & 		   &  		   &  		   &  		    & 		   &  		    & 1.00  & $-$0.01  \\ 
$\langle A_9 \rangle$&           &             &               &             &              &            &              & 		    & 		    & 		   &  		   &  		   &  		    & 		   &  		    & 		    & 1.00  \\ 

\bottomrule
\end{tabular}}
\end{table}

\begin{table}[ht]
\centering 
\caption{Correlation matrix for the observables $\Acp$, $\langle S_i \rangle$ and $\langle A_i \rangle$ for \Dppmm decays measured in the interval \mbox{$1020 \mevcc< m(\mu^+\mu^-) <1100 \mevcc$}.}
\label{table:corr_pipi_S_A_bin6}
 \resizebox{1.0\textwidth}{!}{  
\begin{tabular}[c]{crrrrrrrrrrrrrrrrrr}
\toprule
  	&    $\Acp$  &  $\langle S_2 \rangle$&$\langle S_3 \rangle$&$\langle S_4 \rangle$&$\langle S_5 \rangle$&$\langle S_6 \rangle$&$\langle S_7 \rangle$&$\langle S_8 \rangle$  & $\langle S_9 \rangle$ &  $\langle A_2 \rangle$&$\langle A_3 \rangle$&$\langle A_4 \rangle$&$\langle A_5 \rangle$&$\langle A_6 \rangle$&$\langle A_7 \rangle$&$\langle A_8 \rangle$  & $\langle A_9 \rangle$ \\
$\Acp$& 1.00 &  0.01 &  0.02 &  $-$0.01 &  0.01  & 0.03 &  $-$0.00  & 0.04  & 0.04 &  $-$0.03 &  $-$0.04 &  0.00 &  $-$0.01  & $-$0.07 &  0.02  & $-$0.09  & $-$0.01  \\ 
$\langle S_2 \rangle$&           &  1.00 &  $-$0.10 &  $-$0.05 &  0.02  & 0.01 &  0.11  & 0.03  & $-$0.03  & $-$0.09 &  0.01 &  0.04 &  $-$0.05  & $-$0.04 &  0.01  & 0.07  & $-$0.09  \\ 
$\langle S_3 \rangle$&           &             &  	 1.00 &  0.03 &  0.00  & $-$0.01 &  $-$0.01  & 0.03  & $-$0.04  & $-$0.07 &  $-$0.07 &  0.01 &  0.03  & 0.05 &  0.09  & 0.03  & 0.05  \\ 
$\langle S_4 \rangle$&           &             &               &  1.00 &  $-$0.02  & $-$0.07 &  $-$0.01  & 0.14  & 0.01  & $-$0.03 &  0.03 &  $-$0.09 &  0.03  & $-$0.04 &  0.03  & $-$0.03  & 0.11  \\ 
$\langle S_5 \rangle$&           &             &               &             &  1.00  & $-$0.19 &  0.11  & 0.01  & 0.03  & $-$0.01 &  $-$0.02 &  0.02 &  $-$0.05  & $-$0.06 &  $-$0.04  & $-$0.03  & $-$0.05  \\ 
$\langle S_6 \rangle$&           &             &               &             &              & 1.00 &  $-$0.05  & $-$0.01  & $-$0.04  & 0.01 &  $-$0.08 &  $-$0.03 &  $-$0.10  & $-$0.05 &  $-$0.04  & $-$0.01  & 0.03  \\ 
$\langle S_7 \rangle$&           &             &               &             &              &            &  1.00  & $-$0.01  & $-$0.07  & $-$0.04 &  0.06 &  0.07 &  $-$0.04  & 0.02 &  0.03  & 0.10  & $-$0.04  \\ 
$\langle S_8 \rangle$&           &             &               &             &              &            &              & 1.00  & $-$0.08  & $-$0.01 &  0.06 &  0.01 &  0.02  & 0.04 &  0.03  & $-$0.06  & 0.02  \\ 
$\langle S_9 \rangle$&           &             &               &             &              &            &              & 		    & 1.00  & $-$0.06 &  0.07 &  0.09 &  0.02  & 0.06 &  $-$0.04  & $-$0.03  & $-$0.08  \\ 
$\langle A_2 \rangle$&           &             &               &             &              &            &              & 		    & 		    & 1.00 &  $-$0.03 &  $-$0.05 &  0.07  & 0.01 &  0.08  & 0.05  & $-$0.02  \\ 
$\langle A_3 \rangle$&           &             &               &             &              &            &              & 		    & 		    & 		   &  1.00 &  0.02 &  0.04  & $-$0.03 &  $-$0.04  & 0.02  & 0.09  \\ 
$\langle A_4 \rangle$&           &             &               &             &              &            &              & 		    & 		    & 		   &  		   &  1.00 &  0.01  & 0.00 &  0.01  & $-$0.01  & $-$0.05  \\ 
$\langle A_5 \rangle$&           &             &               &             &              &            &              & 		    & 		    & 		   &  		   &  		   &  1.00  & $-$0.10 &  0.10  & $-$0.01  & $-$0.03  \\ 
$\langle A_6 \rangle$&           &             &               &             &              &            &              & 		    & 		    & 		   &  		   &  		   &  		    & 1.00 &  $-$0.01  & 0.04  & 0.04  \\ 
$\langle A_7 \rangle$&           &             &               &             &              &            &              & 		    & 		    & 		   &  		   &  		   &  		    & 		   &  1.00  & 0.01  & $-$0.04  \\ 
$\langle A_8 \rangle$&           &             &               &             &              &            &              & 		    & 		    & 		   &  		   &  		   &  		    & 		   &  		    & 1.00  & $-$0.12  \\ 
$\langle A_9 \rangle$&           &             &               &             &              &            &              & 		    & 		    & 		   &  		   &  		   &  		    & 		   &  		    & 		    & 1.00  \\ 
\bottomrule
\end{tabular}}
\end{table}

\begin{table}[ht]
\centering 
\caption{Correlation matrix for the observables $\Acp$, $\langle S_i \rangle$ and $\langle A_i \rangle$ for \Dkkmm decays measured in the dimuon-mass-integrated interval.}\label{table:corr_KK_S_A_bin0}
 \resizebox{1.0\textwidth}{!}{  
\begin{tabular}[c]{crrrrrrrrrrrrrrrrrr}
\toprule
  	&    $\Acp$  &  $\langle S_2 \rangle$&$\langle S_3 \rangle$&$\langle S_4 \rangle$&$\langle S_5 \rangle$&$\langle S_6 \rangle$&$\langle S_7 \rangle$&$\langle S_8 \rangle$  & $\langle S_9 \rangle$ &  $\langle A_2 \rangle$&$\langle A_3 \rangle$&$\langle A_4 \rangle$&$\langle A_5 \rangle$&$\langle A_6 \rangle$&$\langle A_7 \rangle$&$\langle A_8 \rangle$  & $\langle A_9 \rangle$ \\

$\Acp$& 1.00 &  $-$0.01 &  $-$0.03 &  0.03 &  0.00  & $-$0.02 &  $-$0.03  & 0.02  & $-$0.02 &  0.08 &  0.03 &  0.00 &  0.01  & 0.02 &  $-$0.00  & 0.03  & $-$0.06  \\ 
$\langle S_2 \rangle$&           &  1.00 &  0.08 &  $-$0.09 &  $-$0.12  & 0.08 &  0.13  & 0.03  & $-$0.03  & 0.07 &  $-$0.01 &  $-$0.06 &  $-$0.09  & $-$0.01 &  $-$0.08  & $-$0.08  & $-$0.03  \\ 
$\langle S_3 \rangle$&           &             &  	 1.00 &  $-$0.08 &  0.05  & 0.05 &  $-$0.05  & $-$0.01  & 0.05  & $-$0.02 &  0.05 &  $-$0.03 &  $-$0.00  & $-$0.08 &  $-$0.10  & $-$0.02  & 0.09  \\ 
$\langle S_4 \rangle$&           &             &               &  1.00 &  0.07  & $-$0.13 &  $-$0.03  & 0.05  & 0.09  & $-$0.07 &  0.00 &  0.01 &  $-$0.01  & 0.03 &  0.04  & 0.06  & $-$0.06  \\ 
$\langle S_5 \rangle$&           &             &               &             &  1.00  & $-$0.09 &  0.08  & $-$0.04  & $-$0.01  & $-$0.11 &  $-$0.01 &  $-$0.00 &  0.01  & $-$0.11 &  0.11  & 0.04  & 0.03  \\ 
$\langle S_6 \rangle$&           &             &               &             &              & 1.00 &  0.05  & $-$0.00  & $-$0.03  & $-$0.01 &  $-$0.07 &  0.04 &  $-$0.07  & 0.03 &  $-$0.05  & 0.01  & 0.01  \\ 
$\langle S_7 \rangle$&           &             &               &             &              &            &  1.00  & 0.08  & $-$0.18  & $-$0.08 &  $-$0.09 &  0.03 &  0.06  & $-$0.06 &  0.04  & $-$0.01  & 0.03  \\ 
$\langle S_8 \rangle$&           &             &               &             &              &            &              & 1.00  & $-$0.13  & $-$0.07 &  $-$0.02 &  0.05 &  0.01  & $-$0.00 &  $-$0.01  & 0.03  & $-$0.09  \\ 
$\langle S_9 \rangle$&           &             &               &             &              &            &              & 		    & 1.00  & $-$0.03 &  0.08 &  $-$0.05 &  0.05  & 0.03 &  0.02  & $-$0.07  & 0.01  \\ 
$\langle A_2 \rangle$&           &             &               &             &              &            &              & 		    & 		    & 1.00 &  0.05 &  $-$0.08 &  $-$0.11  & 0.08 &  0.13  & 0.06  & $-$0.04  \\ 
$\langle A_3 \rangle$&           &             &               &             &              &            &              & 		    & 		    & 		   &  1.00 &  $-$0.09 &  0.06  & 0.06 &  $-$0.09  & $-$0.02  & 0.05  \\ 
$\langle A_4 \rangle$&           &             &               &             &              &            &              & 		    & 		    & 		   &  		   &  1.00 &  0.06  & $-$0.14 &  $-$0.04  & 0.05  & 0.08  \\ 
$\langle A_5 \rangle$&           &             &               &             &              &            &              & 		    & 		    & 		   &  		   &  		   &  1.00  & $-$0.11 &  0.05  & $-$0.01  & 0.01  \\ 
$\langle A_6 \rangle$&           &             &               &             &              &            &              & 		    & 		    & 		   &  		   &  		   &  		    & 1.00 &  0.03  & $-$0.01  & $-$0.03  \\ 
$\langle A_7 \rangle$&           &             &               &             &              &            &              & 		    & 		    & 		   &  		   &  		   &  		    & 		   &  1.00  & 0.08  & $-$0.18  \\ 
$\langle A_8 \rangle$&           &             &               &             &              &            &              & 		    & 		    & 		   &  		   &  		   &  		    & 		   &  		    & 1.00  & $-$0.13  \\ 
$\langle A_9 \rangle$&           &             &               &             &              &            &              & 		    & 		    & 		   &  		   &  		   &  		    & 		   &  		    & 		    & 1.00  \\ 
 
\bottomrule
\end{tabular}}
\end{table}

\begin{table}[ht]
\centering 
\caption{Correlation matrix for the observables $\Acp$, $\langle S_i \rangle$ and $\langle A_i \rangle$ for \Dkkmm decays measured in the interval \mbox{$m(\mu^+\mu^-) <525 \mevcc$}.}
\label{table:corr_KK_S_A_bin1}
 \resizebox{1.0\textwidth}{!}{  
\begin{tabular}[c]{crrrrrrrrrrrrrrrrrr}
\toprule
  	&    $\Acp$  &  $\langle S_2 \rangle$&$\langle S_3 \rangle$&$\langle S_4 \rangle$&$\langle S_5 \rangle$&$\langle S_6 \rangle$&$\langle S_7 \rangle$&$\langle S_8 \rangle$  & $\langle S_9 \rangle$ &  $\langle A_2 \rangle$&$\langle A_3 \rangle$&$\langle A_4 \rangle$&$\langle A_5 \rangle$&$\langle A_6 \rangle$&$\langle A_7 \rangle$&$\langle A_8 \rangle$  & $\langle A_9 \rangle$ \\
$\Acp$& 1.00 &  0.07 &  $-$0.09 &  0.07 &  $-$0.09  & $-$0.09 &  $-$0.09  & $-$0.06  & 0.01 &  0.18 &  0.07 &  $-$0.10 &  $-$0.03  & $-$0.03 &  $-$0.08  & 0.06  & $-$0.04  \\ 
$\langle S_2 \rangle$&           &  1.00 &  0.03 &  $-$0.06 &  $-$0.23  & 0.04 &  $-$0.17  & 0.16  & 0.00  & $-$0.02 &  $-$0.07 &  0.15 &  $-$0.19  & $-$0.20 &  $-$0.03  & $-$0.12  & 0.07  \\ 
$\langle S_3 \rangle$&           &             &  	 1.00 &  0.03 &  $-$0.03  & 0.14 &  0.00  & 0.09  & $-$0.17  & $-$0.06 &  $-$0.06 &  0.26 &  0.03  & $-$0.28 &  $-$0.08  & $-$0.04  & 0.12  \\ 
$\langle S_4 \rangle$&           &             &               &  1.00 &  $-$0.04  & $-$0.37 &  $-$0.19  & 0.01  & 0.23  & 0.16 &  0.27 &  $-$0.11 &  0.06  & 0.11 &  0.34  & 0.40  & $-$0.21  \\ 
$\langle S_5 \rangle$&           &             &               &             &  1.00  & 0.11 &  0.04  & $-$0.13  & $-$0.15  & $-$0.19 &  0.03 &  0.06 &  $-$0.07  & $-$0.15 &  0.36  & 0.30  & 0.06  \\ 
$\langle S_6 \rangle$&           &             &               &             &              & 1.00 &  0.13  & $-$0.09  & $-$0.27  & $-$0.21 &  $-$0.29 &  0.09 &  $-$0.13  & $-$0.07 &  $-$0.15  & $-$0.15  & 0.36  \\ 
$\langle S_7 \rangle$&           &             &               &             &              &            &  1.00  & $-$0.09  & $-$0.35  & $-$0.02 &  $-$0.09 &  0.33 &  0.36  & $-$0.15 &  $-$0.13  & 0.15  & 0.19  \\ 
$\langle S_8 \rangle$&           &             &               &             &              &            &              & 1.00  & 0.19  & $-$0.12 &  $-$0.05 &  0.41 &  0.31  & $-$0.13 &  0.13  & $-$0.08  & $-$0.24  \\ 
$\langle S_9 \rangle$&           &             &               &             &              &            &              & 		    & 1.00  & 0.04 &  0.14 &  $-$0.22 &  0.07  & 0.40 &  0.21  & $-$0.24  & $-$0.27  \\ 
$\langle A_2 \rangle$&           &             &               &             &              &            &              & 		    & 		    & 1.00 &  0.02 &  $-$0.06 &  $-$0.25  & 0.03 &  $-$0.16  & 0.18  & 0.00  \\ 
$\langle A_3 \rangle$&           &             &               &             &              &            &              & 		    & 		    & 		   &  1.00 &  0.04 &  $-$0.03  & 0.14 &  $-$0.00  & 0.10  & $-$0.12  \\ 
$\langle A_4 \rangle$&           &             &               &             &              &            &              & 		    & 		    & 		   &  		   &  1.00 &  $-$0.03  & $-$0.34 &  $-$0.20  & 0.04  & 0.21  \\ 
$\langle A_5 \rangle$&           &             &               &             &              &            &              & 		    & 		    & 		   &  		   &  		   &  1.00  & 0.10 &  0.06  & $-$0.14  & $-$0.14  \\ 
$\langle A_6 \rangle$&           &             &               &             &              &            &              & 		    & 		    & 		   &  		   &  		   &  		    & 1.00 &  0.14  & $-$0.10  & $-$0.25  \\ 
$\langle A_7 \rangle$&           &             &               &             &              &            &              & 		    & 		    & 		   &  		   &  		   &  		    & 		   &  1.00  & $-$0.10  & $-$0.34  \\ 
$\langle A_8 \rangle$&           &             &               &             &              &            &              & 		    & 		    & 		   &  		   &  		   &  		    & 		   &  		    & 1.00  & 0.19  \\ 
$\langle A_9 \rangle$&           &             &               &             &              &            &              & 		    & 		    & 		   &  		   &  		   &  		    & 		   &  		    & 		    & 1.00  \\ 
\hline\hline
\end{tabular}}
\end{table}

\begin{table}[ht]
\centering 
\caption{Correlation matrix for the observables $\Acp$, $\langle S_i \rangle$ and $\langle A_i \rangle$ for \Dkkmm decays measured in the interval \mbox{$m(\mu^+\mu^-) >565 \mevcc$}.}
\label{table:corr_KK_S_A_bin3}
 \resizebox{1.0\textwidth}{!}{  
\begin{tabular}[c]{crrrrrrrrrrrrrrrrrr}
\toprule
  	&    $\Acp$  &  $\langle S_2 \rangle$&$\langle S_3 \rangle$&$\langle S_4 \rangle$&$\langle S_5 \rangle$&$\langle S_6 \rangle$&$\langle S_7 \rangle$&$\langle S_8 \rangle$  & $\langle S_9 \rangle$ &  $\langle A_2 \rangle$&$\langle A_3 \rangle$&$\langle A_4 \rangle$&$\langle A_5 \rangle$&$\langle A_6 \rangle$&$\langle A_7 \rangle$&$\langle A_8 \rangle$  & $\langle A_9 \rangle$ \\
$\Acp$& 1.00 &  $-$0.02 &  $-$0.02 &  $-$0.01 &  $-$0.01  & $-$0.00 &  0.00  & 0.01  & $-$0.01 &  0.06 &  0.02 &  0.01 &  $-$0.02  & 0.03 &  $-$0.01  & $-$0.00  & $-$0.04  \\ 
$\langle S_2 \rangle$&           &  1.00 &  0.06 &  $-$0.09 &  $-$0.05  & 0.08 &  0.18  & 0.02  & $-$0.05  & 0.04 &  0.04 &  $-$0.10 &  $-$0.11  & 0.02 &  $-$0.11  & $-$0.06  & $-$0.04  \\ 
$\langle S_3 \rangle$&           &             &  	 1.00 &  $-$0.12 &  0.10  & 0.04 &  $-$0.09  & $-$0.03  & 0.08  & 0.01 &  0.09 &  $-$0.13 &  $-$0.01  & $-$0.01 &  $-$0.08  & 0.02  & 0.06  \\ 
$\langle S_4 \rangle$&           &             &               &  1.00 &  0.08  & $-$0.09 &  0.02  & 0.05  & 0.03  & $-$0.11 &  $-$0.13 &  0.03 &  $-$0.02  & 0.01 &  $-$0.02  & 0.01  & $-$0.05  \\ 
$\langle S_5 \rangle$&           &             &               &             &  1.00  & $-$0.16 &  0.05  & 0.00  & 0.03  & $-$0.09 &  $-$0.02 &  $-$0.01 &  0.06  & $-$0.08 &  0.00  & $-$0.05  & 0.05  \\ 
$\langle S_6 \rangle$&           &             &               &             &              & 1.00 &  0.03  & 0.03  & 0.01  & 0.03 &  0.02 &  $-$0.01 &  $-$0.11  & 0.06 &  $-$0.03  & 0.06  & $-$0.05  \\ 
$\langle S_7 \rangle$&           &             &               &             &              &            &  1.00  & 0.09  & $-$0.11  & $-$0.08 &  $-$0.08 &  $-$0.04 &  0.01  & $-$0.05 &  0.05  & $-$0.02  & $-$0.00  \\ 
$\langle S_8 \rangle$&           &             &               &             &              &            &              & 1.00  & $-$0.15  & $-$0.07 &  0.01 &  0.01 &  $-$0.04  & 0.06 &  $-$0.03  & 0.03  & $-$0.07  \\ 
$\langle S_9 \rangle$&           &             &               &             &              &            &              & 		    & 1.00  & $-$0.06 &  0.08 &  $-$0.05 &  0.05  & $-$0.07 &  $-$0.00  & $-$0.07  & 0.05  \\ 
$\langle A_2 \rangle$&           &             &               &             &              &            &              & 		    & 		    & 1.00 &  0.06 &  $-$0.11 &  $-$0.06  & 0.10 &  0.19  & 0.04  & $-$0.03  \\ 
$\langle A_3 \rangle$&           &             &               &             &              &            &              & 		    & 		    & 		   &  1.00 &  $-$0.12 &  0.10  & 0.01 &  $-$0.09  & $-$0.02  & 0.06  \\ 
$\langle A_4 \rangle$&           &             &               &             &              &            &              & 		    & 		    & 		   &  		   &  1.00 &  0.09  & $-$0.09 &  0.01  & 0.07  & 0.04  \\ 
$\langle A_5 \rangle$&           &             &               &             &              &            &              & 		    & 		    & 		   &  		   &  		   &  1.00  & $-$0.16 &  0.06  & $-$0.01  & 0.05  \\ 
$\langle A_6 \rangle$&           &             &               &             &              &            &              & 		    & 		    & 		   &  		   &  		   &  		    & 1.00 &  0.05  & 0.04  & $-$0.02  \\ 
$\langle A_7 \rangle$&           &             &               &             &              &            &              & 		    & 		    & 		   &  		   &  		   &  		    & 		   &  1.00  & 0.09  & $-$0.09  \\ 
$\langle A_8 \rangle$&           &             &               &             &              &            &              & 		    & 		    & 		   &  		   &  		   &  		    & 		   &  		    & 1.00  & $-$0.14  \\ 
$\langle A_9 \rangle$&           &             &               &             &              &            &              & 		    & 		    & 		   &  		   &  		   &  		    & 		   &  		    & 		    & 1.00  \\ 
\bottomrule
\end{tabular}}
\end{table}

\clearpage 
 
\newpage
%
%
%
\centerline
{\large\bf LHCb collaboration}
\begin
{flushleft}
\small
R.~Aaij$^{32}$,
A.S.W.~Abdelmotteleb$^{56}$,
C.~Abell{\'a}n~Beteta$^{50}$,
F.~Abudin{\'e}n$^{56}$,
T.~Ackernley$^{60}$,
B.~Adeva$^{46}$,
M.~Adinolfi$^{54}$,
H.~Afsharnia$^{9}$,
C.~Agapopoulou$^{13}$,
C.A.~Aidala$^{87}$
% ...

{\footnotesize \it

% ...
$^{9}$Universit{\'e} Clermont Auvergne, CNRS/IN2P3, LPC, Clermont-Ferrand, France\\
% ...
$^{13}$LPNHE, Sorbonne Universit{\'e}, Paris Diderot Sorbonne Paris Cit{\'e}, CNRS/IN2P3, Paris, France\\
% ...
$^{32}$Nikhef National Institute for Subatomic Physics, Amsterdam, Netherlands\\
% ...
$^{46}$Instituto Galego de F{\'\i}sica de Altas Enerx{\'\i}as (IGFAE), Universidade de Santiago de Compostela, Santiago de Compostela, Spain\\
% ...
$^{50}$Physik-Institut, Universit{\"a}t Z{\"u}rich, Z{\"u}rich, Switzerland\\
% ...
$^{54}$H.H. Wills Physics Laboratory, University of Bristol, Bristol, United Kingdom\\
% ...
$^{56}$Department of Physics, University of Warwick, Coventry, United Kingdom\\
% ...
$^{60}$Oliver Lodge Laboratory, University of Liverpool, Liverpool, United Kingdom\\
% ...
$^{68}$Syracuse University, Syracuse, NY, United States\\
% ...
$^{87}$University of Michigan, Ann Arbor, United States, associated to $^{68}$\\
\bigskip
% ...
}
\end{flushleft} 
 
\end{document}